\documentclass[mathpazo]{cicp}

\usepackage{stmaryrd,amsmath,amssymb}
\usepackage{tabularx,epsfig,color,tabularx}
\usepackage{algorithm,algorithmic,multirow}
\usepackage{graphicx,caption,subcaption}
\usepackage{mathtools,epstopdf,hyperref}

\input psfig.sty

\newcommand{\gm}{\gamma}

\newcommand\bx{\mathbf{x}}
\newcommand\bn{\mathbf{n}}
\newcommand\bm{\mathbf{m}}

\newcommand\bk{\mathbf{k}}

\newcommand\by{\mathbf{y}}
\newcommand\bR{\mathbb{R}}

\newcommand\wrho{\widehat{\rho}}

\newcommand{\p}{\partial}

\newcommand{\Og}{\Omega}
\newcommand{\fl}[2]{\frac{#1}{#2}}

\newcommand{\nn}{\nonumber}

\newcommand{\bt}{\beta}

\newcommand{\Dt}{\Delta}

\newcommand{\be}{\begin{equation}}
\newcommand{\ee}{\end{equation}}
\newcommand{\ba}{\begin{array}}
\newcommand{\ea}{\end{array}}
\def\bea{\begin{eqnarray}}
\def\eea{\end{eqnarray}}
\def \beas{\begin{eqnarray*}}
\def \eeas{\end{eqnarray*}}
\newtheorem{exmp}{Example}[section]

\begin{document}

%\begin{frontmatter}
%
\title{Accurate and efficient numerical methods for computing ground states and dynamics
of dipolar Bose-Einstein condensates via the nonuniform FFT}

 \author[Weizhu Bao et.~al.]{Weizhu Bao\affil{1},
          Qinglin Tang\affil{2}\affil{3}\affil{4}\corrauth and Yong Zhang\affil{5}}
\address{\affilnum{1}\ Department of Mathematics, National University of
Singapore, Singapore 119076\\
\affilnum{2}\ Universit\'e de Lorraine, Institut Elie Cartan de
Lorraine, UMR 7502, Vandoeuvre-l\`es-Nancy, F-54506, France\\
\affilnum{3}\ Inria Nancy Grand-Est/IECL-CORIDA, France\\
\affilnum{4}\ Beijing Computational Science Research Center, Beijing 100084, P. R. China\\
\affilnum{5}\ Wolfgang Pauli Institute c/o Fak. Mathematik,
University Wien, Oskar-Morgenstern-Platz 1, 1090 Vienna, Austria}

\emails{{\tt matbaowz@nus.edu.sg} (W.~Bao), {\tt qinglin.tang@inria.fr} (Q.~Tang),
{\tt yong.zhang@univie.ac.at} (Y.~Zhang)}

%%%%% Begin Abstract %%%%%%%%%%%
\begin{abstract}
In this paper, we propose efficient and accurate numerical methods for computing the ground state
and dynamics of the dipolar Bose-Einstein condensates utilising a newly developed
dipole-dipole interaction (DDI) solver that is implemented with
the non-uniform fast Fourier transform (NUFFT) algorithm.
We begin with the three-dimensional (3D) Gross-Pitaevskii equation (GPE)
with a DDI term and present the corresponding two-dimensional (2D) model
under a strongly anisotropic confining potential. Different from existing methods,
the NUFFT based DDI solver removes the singularity by adopting
the spherical/polar coordinates in Fourier space in 3D/2D, respectively,
thus it can achieve spectral accuracy in space and simultaneously maintain
high efficiency   by making full use of
FFT and NUFFT whenever it is necessary and/or needed.
Then, we incorporate this solver into existing successful methods
for computing the ground state and dynamics of GPE with a DDI for dipolar BEC.
Extensive numerical comparisons with  existing methods are carried out
for computing the DDI, ground states and dynamics of the dipolar BEC.
Numerical results show that our new methods outperform existing methods
in terms of both accuracy and efficiency.
\end{abstract}


\maketitle{}

\begin{center}
{ \bf Dedicated to Professor Eitan Tadmor on the occasion of his 60th birthday}
\end{center}

%%%%%%%%%%%%%%%%%%%%%%%%%%%%%%%%%%%
\section{Introduction}
Since its first experimental creation in 1995 \cite{AEMWC1995, BSTH1995, DMADDKK1995},
the Bose-Einstein condensation (BEC) has provided an incredible glimpse into the macroscopic quantum world and
opened a new era in atomic and molecular physics as well as condensated matter physics.
It regains vast interests and has been extensively studied both experimentally and theoretically \cite{A2004, Bar2008,
BDZ2008, F2009, L2001,  MO2006, PS2003}.   At early stage, experiments mainly realize
BECs of ultracold atomic gases whose properties are mainly governed by the isotropic and short-range interatomic interactions \cite{PS2003}.
However, recent experimental developments on Feshbach resonances \cite{Lah2007}, on cooling and trapping molecules
\cite{Ni2008, Shu2010} and on precision measurements and control \cite{Ven2008, Pol2009} allow one to realize BECs of
quantum gases with different, richer interactions and gain even more interesting properties.  In particular, the successful realization
of BECs of dipolar quantum gases with long-range and anisotropic  dipolar interaction,
e.g., $^{52}{\rm Cr}$ \cite{Gri2005}, $^{164}{\rm Dy}$
\cite{Lu2011} and $^{168}{\rm Er}$ \cite{Aik2012}, has spurred great interests in the unique properties of
degenerate dipolar quantum gases
and stimulated enthusiasm in studying both the ground state \cite{ DipJCP, BC2013, Jiang2006, Santos2000,  YY2001}
and dynamics \cite{BMTZ2013, BJM2004, CMS2008, HMS2010, Lah2008, Park2009} of dipolar BECs.

At temperatures $T$ much smaller than the critical  temperature $T_c$,  the properties of
BEC with long-range dipole-dipole interactions (DDI) are well described by the
macroscopic  complex-valued wave function $\psi=\psi(\bx,t)$ whose evolution
is governed by the celebrating three-dimensional (3D) Gross--Pitaevskii equation (GPE) with a DDI term.
Moreover,  the 3D GPE can be reduced to an effective two-dimensional (2D) version if the external
trapping potential is strongly confined
 in the $z-$direction \cite{CRLB2010, DipJCP}. In a unified way,
 the dimensionless
 GPE with a DDI term in $d-$dimensions ($d=2\ {\rm or}\ 3$) for modeling a dipolar BEC
 reads as \cite{BMTZ2013, BAC2012, BC2013, GRP2000,  YY2001}:
\bea\label{DipGPE0}
&& i\p_t \psi({\bx}, t) = \left[-\fl{1}{2}\nabla^2 + V({\bf x}) + \beta |\psi|^2 +
\lambda\, \Phi(\bx,t) \right]\psi(\bx,t),\quad \bx\in{\mathbb R}^d, \quad t>0, \\
\label{dipole-poten0}
&&\Phi(\bx,t)=\left(U_{\rm dip}\ast |\psi|^2\right)(\bx,t), \qquad\qquad \bx\in{\mathbb R}^d, \quad t\ge0,\\
\label{ini-con0}
&&\psi(\bx,t=0)=\psi_0(\bx), \qquad\qquad\; \bx\in  {\mathbb R}^d,
\eea
where $t$ is time, ${\bf x}=(x, y)^T \in {\mathbb R}^2$ or ${\bf x}=(x, y, z)^T \in {\mathbb R}^3$,
$\ast$ represents the convolution operator with respect to spatial variable.
The dimensionless constant $\beta$ describes the strength of the short-range two-body
interactions  in a condensate (positive for repulsive
interaction, and resp. negative for attractive interaction), while
$V(\bx)$ is a given real-valued external trapping potential
which is determined by the type of system under investigation.  In most BEC experiments,
a harmonic potential is chosen to trap the condensate, i.e.,\
\be\label{Vpoten}
V(\bx) = \fl{1}{2}\left\{\begin{array}{ll}
 \gm_x^2x^2 + \gm_y^2y^2, & d = 2,\\[0.3em]
\gm_x^2x^2 + \gm_y^2y^2 + \gm_z^2z^2, \ \ &d = 3,
\end{array}\right.
\ee
where $\gm_x>0$, $\gm_y>0$ and $\gm_z>0$ are  dimensionless constants proportional to the
trapping frequencies in $x$-, $y$- and $z$-direction, respectively.
Moreover, $\lambda$ is a dimensionless constant characterizing the strength of DDI and
$\Phi(\bx,t)$ is the long-range dipole interaction whose convolution kernel in 3D/2D is given as
\cite{BAC2012, BMTZ2013, CRLB2010, DipJCP, BaoJiangLeslie}:
\be\label{DipKer}
U_{\rm dip}(\bx)=
\left\{\begin{array}{l}
-\delta(\bx)-3\,\partial_{\bn\bn}   \left( \fl{1}{4\pi|\bx|} \right), \\[0.5em]
-\fl{3}{2} \left(\partial_{\bn_{\perp}\bn_{\perp}}-n_3^2 \nabla_{\perp}^2 \right)
\left( \fl{1}{2\pi|\bx|} \right),
\end{array}
\right.
\Leftrightarrow\; \;
\widehat{U}_{\rm dip} (\bk)= \left\{\begin{array}{ll}
		-1 + \fl{3 (\bn\cdot\bk)^2}{\Vert\bk\Vert^2},   & d=3, \\[0.5em]
\fl{3\left[ (\bn_{\perp}\cdot\bk)^2-n_3^2 \Vert\bk\Vert^2 \right]}{2\Vert\bk\Vert}, &   d=2,
\end{array}\right.
\ee
where $\bx,\bk\in \mathbb{R}^d$ and $\widehat{f}(\bk) =  \int_{{\mathbb R}^d} f(\bx)\;e^{-i \bk\cdot \bx}\, d\bx$
is the Fourier transform of $f(\bx)$.
Here, $\bn = (n_1, n_2, n_3)^T$ is a given  unit vector
representing the dipole axis, $\bn_\perp=(n_1,n_2)^T$,
 $\partial_\bn=\bn\cdot \nabla$,
 $\partial_{\bn\bn}=\partial_\bn(\partial_\bn)$,
 $\nabla_\perp=(\partial_x, \partial_y)^T$,
 $\partial_{\bn_\perp}=\bn_\perp \cdot \nabla_\perp$ and
$\partial_{\bn_{\perp}\bn_{\perp}}=\partial_{\bn_{\perp}}(\partial_{\bn_{\perp}}).$
Note that the dipole axis can also be
different. The dipole kernel $U_{\rm dip}(\bx)$ with  two different dipole orientations  $\bn$ and $\bm$
reads as \cite{BaoJiangLeslie, ODell2009, BAC2012}
\be
\label{dipmn}
U_{\rm dip}(\bx)=
\left\{\begin{array}{ll}
 -(\bn\cdot\bm) \delta(\bx)-3\,\partial_{\bn\bm}   \left( \fl{1}{4\pi|\bx|} \right),  &  \bx\in{\Bbb R}^3,\\
-\fl{3}{2} \left(\partial_{\bn_{\perp}\bm_{\perp}}-n_3 m_3 \nabla_{\perp}^2 \right) \left( \fl{1}{2\pi|\bx|} \right),   &  \bx\in{\Bbb R}^2,
\end{array}
\right.
\ee
where $\bm = (m_1, m_2, m_3)^T$ is a given  unit vector representing the other dipole orientation, $\bm_\perp=(m_1,m_2)^T$,
 $\partial_{\bm_\perp}=\bm_\perp \cdot \nabla_\perp$ and
$\partial_{\bn_{\perp}\bm_{\perp}}=\partial_{\bn_{\perp}}(\partial_{\bm_{\perp}})$.
We remark here that in most physical experiments, the dipoles are polarized  at the same direction,
i.e., $\bm=\bn$, thus, hereafter, we always assume $\bm=\bn$ unless specified otherwise.

The GPE (\ref{DipGPE0})-(\ref{ini-con0}) conserves two important quantities:  the
{\it mass} (or {\it normalization}) of the wave function
\bea\label{norm}
N(t):=\|\psi(\cdot, t)\|^2 := \int_{{\Bbb R}^d}|\psi({\bx}, t)|^2 d {\bf x}
\equiv \int_{{\Bbb R}^d}|\psi({\bf x}, 0)|^2 d {\bf x} = 1, \qquad t\geq 0,
\eea
and the {\it energy  per particle}
\be\label{energy}
E(\psi(\cdot, t))=\int_{{\Bbb R}^d}\left[\fl{1}{2}|\nabla\psi|^2 + V({\bf x})|\psi|^2+\fl{\bt}{2}
|\psi|^4+\fl{\lambda}{2}\Phi(\bx,t)\, |\psi|^2\right] d  {\bf x}
\equiv E(\psi(\cdot, 0)), \qquad t\geq 0.
\ee
The ground state $\phi_g$ of the GPE  (\ref{DipGPE0})-(\ref{ini-con0})
is  defined as follows:
\be\label{ground}
\phi_g =\arg \min_{\phi\in S} E(\phi), \quad \hbox{where}\quad  S:=\{\phi(\bx) \ |\
\|\phi\|^2:=\int_{{\mathbb R}^d} |\phi(\bx)|^2d\bx=1, \ E(\phi)<\infty\}.
\ee
Extensive works have been carried out to study the ground state and dynamics of dipolar BEC based on the GPE (\ref{DipGPE0})-(\ref{ini-con0}). For existing theoretical and numerical studies, we refer to  \cite{CMS2008,  BC2013, Bar2008, LMSLP2009, CRLB2010, KU2012, Lah2008} and \cite{BJM2003,BJM2004, Bla2009, HMS2010, GRP2000, Lah2008,   Tick2008,ABB2013}, respectively, and references therein.

To compute the ground state and dynamics of the GPE  (\ref{DipGPE0}), one of the key difficulties is
how to evaluate the nonlocal dipole interaction  $\Phi(\bx,t)$ (\ref{dipole-poten0}) accurately and effectively for a given
density $\rho=|\psi|^2$.   Noticing that
\be
\label{DDI-Con}
\Phi(\bx,t)= \int _{{\Bbb R}^d} U_{\rm dip}(\bx-\by) \rho(\by,t) d\by  =\fl{1}{(2\pi)^d} \int _{{\Bbb R}^d}
\widehat{U}_{\rm dip}(\bk)\, \widehat{\rho}(\bk,t) \,e^{i\, \bk\cdot\bx}\, d\bk,
\ee
it is natural to evaluate $\Phi(\bx,t)$ via the standard fast Fourier transform (FFT)
using a uniform grid on a bounded computational domain \cite{Bla2009, ODell2009, Park2009, Tick2008}.
Nevertheless,  due to the intrinsic singularity/discontinuity
of $\widehat{U}_{\rm dip}(\bk)$ at the origin $\bk={\bf 0}$,
the so called ``numerical locking" phenomena occurs,
which limits the optimal accuracy on any given computational domain
\cite{BC2013,SPMCompare}.
To alleviate this problem, another approach \cite{DipJCP, CRLB2010} is to reformulate the convolution (\ref{dipole-poten0})
with 3D dipole kernel (\ref{DipKer}) in terms of the Poisson equation:
\be
\label{Laplace}
-\Dt\; u(\bx, t) =|\psi(\bx, t)|^2,  \qquad \lim_{|\bx|\rightarrow \infty } u(\bx, t)=0, \qquad  \bx\in{\Bbb R}^3,    \quad t\ge0,
\ee
and convolution (\ref{dipole-poten0})  with 2D dipole kernel (\ref{DipKer}) in terms of the fractional Position equation
\be
\label{Sqare-Lap}
\sqrt{-\Dt}\;  u(\bx, t) =|\psi(\bx, t)|^2,	\qquad	\lim_{|\bx|\rightarrow \infty } u(\bx,t)	=0, \qquad  \bx\in{\Bbb R}^2,    \quad t\ge0.
\ee
Then, the dipolar potential $ \Phi(\bx,t)$  can be computed  by a differentiation of $u(\bx,t)$ as:
\be
\label{DDI-Eqv}
  \Phi(\bx,t)=
\left\{\begin{array}{ll}
-|\psi(\bx, t)|^2-3 \partial_{\bn\bn} u(\bx, t),  &  \bx\in{\Bbb R}^3,\\[0.8em]
-\fl{3}{2}\left(\partial_{\bn_{\perp}\bn_{\perp}}-n_3^2\nabla_{\perp}^2 \right) u(\bx,t),   &  \bx\in{\Bbb R}^2,
\end{array}
\right.
\qquad t\ge0.
\ee
Then in practical computations, the sine pseudospectral method is
applied  to solve (\ref{Laplace})-(\ref{DDI-Eqv}) on a truncated
rectangular domain $\Og$ with homogeneous Dirichlet boundary conditions imposed on $\p\Og$, and
they can be implemented with discrete sine transform (DST)  efficiently and accurately \cite{DipJCP}.
By waiving the
use of the ${\bf 0}$-mode in the Fourier space, the sine spectral method significantly improves the accuracy
for the dipole interaction evaluation.
However, due to the polynomial decaying property of $u(\bx, t)$ when $|\bx|\to\infty$, a very large computational
domain is required in order to achieve satisfactory accuracy.
This will increase the computational cost and storage significantly for the dipole interaction evaluation
and hence for computing the ground state and dynamics of the GPE  (\ref{DipGPE0}).
Moreover, we  shall also remark here that, in most applications, a much smaller domain suffices
the GPE  (\ref{DipGPE0}) simulation because of the exponential decay property of the wave function $\psi(\bx,t)$.

Recently,  an accurate and fast algorithm based on the NUFFT algorithm was proposed
for the evaluation of the dipole interaction in 3D/2D \cite{BaoJiangLeslie}.  The method
also evaluates the dipole interaction in the Fourier domain, i.e., via the integral (\ref{DDI-Con}).
Unlike the standard FFT method, by an adoption of spherical/polar coordinates in the Fourier domain
in 3D/2D,  the  singularity/discontinuity of $\widehat{U}_{\rm dip}(\bk)$ at the origin in the integral (\ref{DDI-Con})
is canceled out by the Jacobian introduced by the coordinates transformation.
The integral is then discretized  by a high-order quadrature and the resulted discrete summation is accelerated via the NUFFT algorithm.
The algorithm has $O(N\log N)$ complexity with $N$ being the total number of unknowns in the physical space
and achieves very high accuracy for the dipole interaction evaluation.
The main objectives of this paper are threefold:
(i) to compare numerically the newly developed NUFFT based method with the
existing methods that are based on DST for the evaluation
of these  nonlocal interactions in terms of the size of the computational domain
$\Omega$ and the mesh size of partitioning $\Omega$;
(ii) to propose efficient and accurate numerical methods for the
ground state  computation and dynamics simulation of the GPE with the nonlocal interactions
(\ref{DipGPE0})-(\ref{dipole-poten0})  by
incorporating the NUFFT based nonlocal interaction evaluation algorithm
into the normalized gradient flow method
and the time-splitting Fourier pseudospectral method, respectively,
and (iii)  to test the performance of the methods and apply them to
compute some interesting phenomena.

The paper is organized as follows. In Section 2, we shall briefly review the NUFFT based algorithm
in \cite{BaoJiangLeslie} for the evaluation of the dipole interaction in 3D/2D.
 In Section 3, an efficient and accurate numerical method will be proposed to
compute the ground state of the GPE (\ref{DipGPE0})-(\ref{dipole-poten0})  by
coupling the NUFFT based algorithm for the evaluation of the dipole interaction
and the discrete normalized gradient flow method.
In Section 4, we will present an efficient and accurate numerical method for computing
the dynamics of the GPE (\ref{DipGPE0})-(\ref{dipole-poten0})  by
coupling the NUFFT based algorithm for the evaluation of the dipole interaction
and the time-splitting Fourier pseudospectral method.
Finally, some concluding remarks will be drawn in Section 5.

%%%%%%%%%%%%%%%%%%%%%%%%%%%%%%%%%%%%%%%%%%%%%%
%%%%%%%%%%%%%%%%%%%%%%%%%%%%%%%%%%%%%%%%%%%%%%
\section{Evaluation of the dipole interaction via NUFFT  }
\label{DDINUFFT}

In this section, we will first briefly review the NUFFT based method in \cite{BaoJiangLeslie}
for computing the dipole interaction in 3D/2D, and then compare this method with the existing
DST-based method.

\subsection{NUFFT based algorithm}

Due to the external trapping potential, the solution of the
GPE (\ref{DipGPE0})-(\ref{ini-con0}) will decay exponentially.
Thus, without loss of generality, it is reasonable to assume that
the density $\rho(\bx,t)$  is smooth and decays rapidly,
hence $\widehat{\rho}(\bk,t)$ is also smooth and decays fast.   Therefore,  up to any prescribed
precision $\varepsilon_0$ (e.g.,  $\varepsilon_0=10^{-12}$),  we can respectively choose bounded domains
$\mathcal{D}$ and $B_{R}(0)=:\{|\bk|\le R, \bk\in{\mathbb R}^d\}$ large enough in the physical space and phase space such that
the truncation error of  $\rho(\bx,t)$ and $\widehat{\rho}(\bk,t)$ is negligible.  Note that the convolution only acts
on the spatial variable, to simplify our presentation, hereafter we
omit the temporal variable $t$ and simplify the notation as $\Phi(\bx,t)  \rightarrow \Phi(\bx)$ and $\rho(\bx,t) \rightarrow \rho(\bx)$.

By truncating the integration domain in  (\ref{DDI-Con}) into a $B_{R}(0)$ and adopting the
spherical/polar coordinates in 3D/2D in the phase (or Fourier) space, we have \cite{BaoJiangLeslie}
 \bea
 \Phi(\bx)
 &=& \nn
 \fl{1}{(2\pi)^d} \int_{{\mathbb R}^d} e^{i\,\bk\cdot\bx} \widehat{U}_{\rm dip}(\bk) \widehat{\rho}(\bk)  d\bk
\;\approx\;  \fl{1}{(2\pi)^d} \int_{B_{R}(0)} e^{i\,\bk\cdot\bx} \widehat{U}_{\rm dip}(\bk) \widehat{\rho}(\bk)  d\bk    \\[0.5em]
 &=&\label{DDI-Trun1}
\fl{1}{(2\pi)^d}  \left\{ \begin{array}{ll}
\int_{0}^{R}\int_{0}^{2\pi} e^{i\,\bk\cdot\bx} |\bk| \widehat{U}_{\rm dip}(\bk)\,
 \widehat{\rho}(\bk)\, d|\bk| d \phi,  & d=2,  \\[0.8em]
\int_{0}^{R}\int_{0}^{\pi}\int_{0}^{2\pi} e^{i\,\bk\cdot\bx}\widehat{U}_{\rm dip}(\bk)\,
 |\bk|^2 \sin{\theta}\;\widehat{\rho}(\bk)\, d|\bk| d\theta d\phi, &d=3.
 \end{array}
 \right.
 \eea
It is easy to see that the singularity/discontinuity of $\widehat{U}_{\rm dip}(\bk)$ at the origin
 is canceled  out by the Jacobian $|\bk|^{d-1}$ and hence the  integrand in the above integral is smooth.
High order quadratures are then applied to further discretize the above integral and
 the resulted summation can be  efficiently evaluated by the NUFFT \cite{BaoJiangLeslie}.
The computational cost of this algorithm is $O(N_1\log N_1)+O(N_2)$, where $N_1$
is the total number of equispaced points in the physical space  $\mathcal{D}$ and $N_2$
is  the number of nonequispaced points in the phase space $B_{R}(0)$.
Roughly speaking, $N_2$ is of the  same order as $N_1$,  however, the constant in front of $O(N_2)$ (e.g.,
$24^d$ for $12$-digit accuracy) is much greater than the constant
in front of $O(N_1\log N_1)$. This makes the algorithm considerably slower than the regular FFT,
especially for three dimensional problems.

\begin{figure}[h!]
\centerline{
\psfig{figure=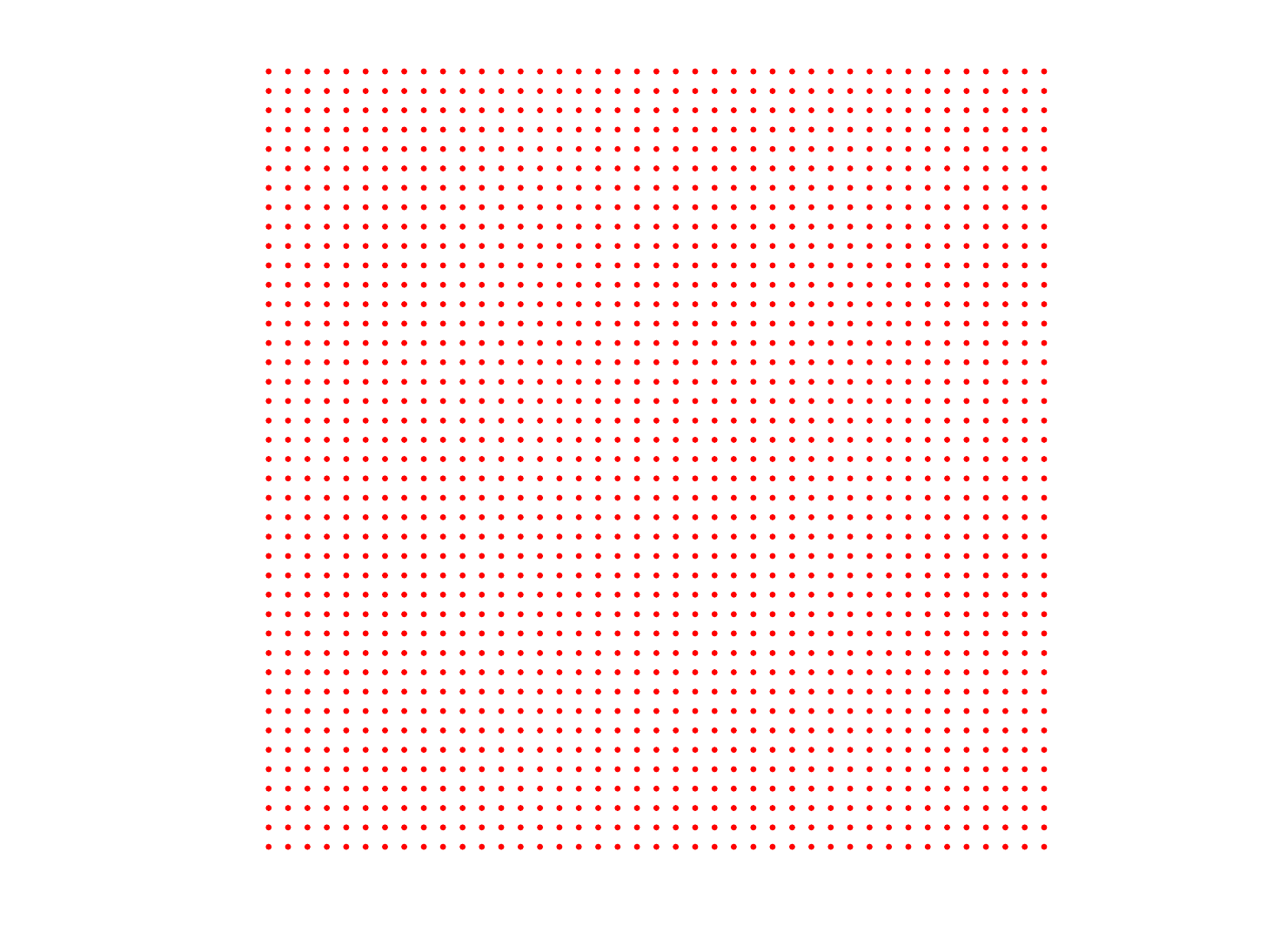,height=5.cm,width=6.8cm,angle=0}\;\;\;\quad
\psfig{figure=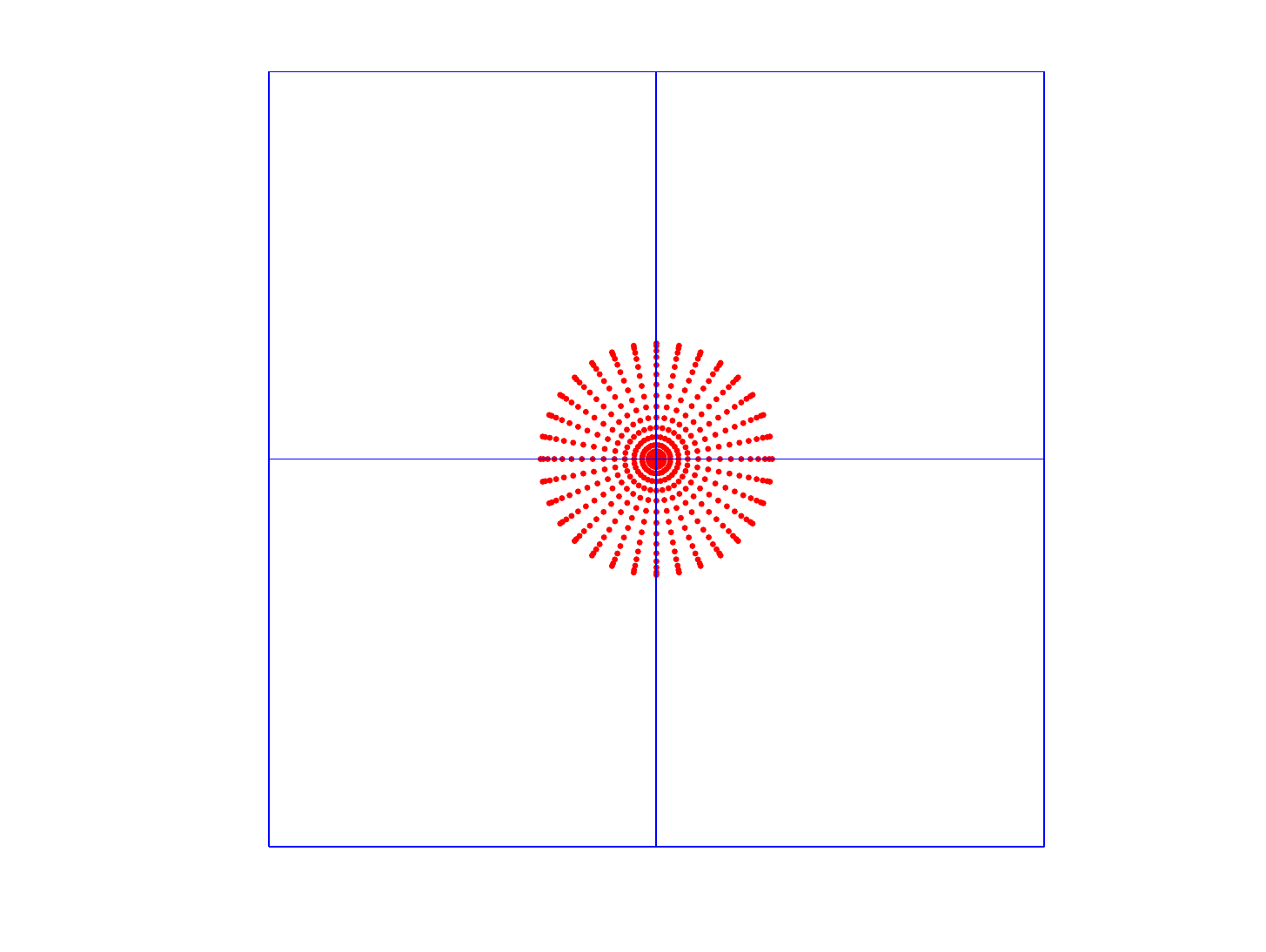,height=5.cm,width=6.8cm,angle=0}\;\;\;\quad
}
\caption{Two grids used in the phase (or Fourier) domain in the improved algorithm in
\cite{BaoJiangLeslie}:
the regular grid on the left panel is used to compute $I_1$ in \eqref{3.9} via the regular FFT;
while the polar grid (confined in a small region centered at the origin)
on the right panel is used to compute
$I_2$ in \eqref{3.9} via the NUFFT. Note that the number of points in the polar grid is
$O(1)$, thus keeping the interpolation cost in NUFFT minimal.
}
\label{fig3.1}
\end{figure}

To reduce the computational cost, an improved algorithm is also proposed  in  \cite{BaoJiangLeslie}.
First, by a simple partition of unity,  the integral in \eqref{DDI-Trun1} can be further  split into two parts:
\bea
\Phi(\bx)
&\approx&\nn
 \fl{1}{(2\pi)^d} \int_{B_{R}(0)}\,e^{i\,\bk\cdot\bx}\,\widehat{U}_{\rm dip}(\bk)\,\widehat{\rho}(\bk) \,d\bk\\
  &=&\nn
 \fl{1}{(2\pi)^d} \int_{B_{R}(0)}\,e^{i\,\bk\cdot\bx}(1-q_d(\bk))\,\widehat{U}_{\rm dip}\,\widehat{\rho}(\bk)\,d\bk
+\fl{1}{(2\pi)^d} \int_{B_{R}(0)}\,e^{i\,\bk\cdot\bx}q_d(\bk)\,\widehat{U}_{\rm dip}\,\widehat{\rho}(\bk)\,d\bk \\
&\approx&\nn
\frac{1}{(2\pi)^{d}}\int_{\Og}\,e^{i\bk\cdot \bx}\,p_d(\bk)\,\wrho(\bk)\,d\bk
+\fl{1}{(2\pi)^d} \int_{B_{R}(0)} e^{i\,\bk\cdot\bx}q_d(\bk) \widehat{U}_{\rm dip}(\bk) \wrho(\bk)  d\bk \\[0.5em]
&:=& \label{3.9}
I_1+I_2,\qquad\qquad\qquad \qquad \bx\in \mathcal{D}.
\eea
Here, $\Og =\{\bk=(k_1,\ldots, k_d)^T \big| |k_l| \leq R, l = 1,\ldots,d\}$ is a rectangular domain
containing the ball $B_{R}(0)$,  the function $q_d(\bk)$ is chosen such that it is a  $C^\infty$ function which
decays exponentially fast as $|\bk|\to\infty$  and the function $p_d(\bk):=(1-q_d(\bk)) \widehat{U}_{\rm dip}(\bk)$
is smooth for $\bk\in\bR^d$.
With this $q_d(\bk)$, $I_1$ can be  evaluated via the regular FFT,
while $I_2$ can be computed via the NUFFT with a fixed (much fewer) number of irregular points
in the phase (or Fourier) space (see Figure \ref{fig3.1}).
Therefore, the interpolation cost in the NUFFT is reduced to $O(1)$
and the overall cost of the  algorithm is comparable to that of the regular FFT,
with a small oversampling factor in front of $O(N_1\log N_1)$.

\subsection{Numerical comparison}

In this subsection, we will show the accuracy and efficiency of the NUFFT based algorithm (referred as {\sl NUFFT})
for computing dipole interaction $\Phi(\bx)$ and compare them with the existing methods that applies (\ref{Laplace})-(\ref{DDI-Eqv})
via the DST (referred as {\sl DST}).  To this end, we denote  $\mathcal{D}$ as  the computational domain, $\mathcal{D}_h$
as its partition with mesh size $h$ and $\Phi_h(\bx)$ as the numerical solution obtained on the domain
$\mathcal{D}_h$. Hereafter, we choose $h_x= h_y=h_z$ in 3D and/or $h_x= h_y$  in 2D and denote them uniformly as $h$
unless stated otherwise.   To demonstrate the comparison, we define the error function as
\be
\label{error_l2}
e_h:=\Vert\Phi-\Phi_{h}\Vert_{l^2}/\Vert \Phi\Vert_{l^2},
\ee
where $\Vert\cdot\Vert_{l^2}$ is the $l^2$-norm.

%%%%%%%%%%%%%%%%%%%%%%%%%%%%%%%%%%%%%%%%%%%%%%
\begin{exmp}\label{exmp_3d} {\sl Dipole-dipole interaction in 3D.}\end{exmp}
In this example, we take $d=3$ and  choose the source density $\rho(\bx)= e^{-|\bx|^{2}/\sigma^{2}}$
with $\sigma>0$.  The 3D dipole interaction with two dipole orientations $\bn$ and $\bm$
can be  given explicitly as
\be
\label{extDensGausDipolar}
 \Phi(\bx) =-(\bn\cdot\bm)\,\rho(\bx) - 3\;
\partial_{\bn\bm} \left(\frac{\sigma^{2}\sqrt{\pi}}{4}  \frac{\text{Erf}(r/\sigma)}{r/\sigma}\right)
= -(\bn\cdot\bm)\,\rho(\bx) - 3\;\bn^{T} G(\bx)\bm,
\ee
where the  matrix $G(\bx)=(g_{jl}(\bx))_{j,l=1}^3$ is given as
\be
\label{Hessian_Ex1}
g_{jl}(\bx)
= \left( \frac{\sigma^{2}}{2r^{2}} e^{-\frac{r^{2}}{\sigma^{2}}} -
\frac{\sigma^{3}\sqrt{\pi}}{4 r^{3}}\text{Erf}\left(\frac{r}{\sigma}\right) \right)\delta_{jl} +
x_{j}x_{l}  \left( -\frac{3 \;\sigma^{2}}{2\;r^{4}} e^{-\frac{r^{2}}{\sigma^{2}}} -
\frac{1}{r^{2}} e^{-\frac{r^{2}}{\sigma^{2}}} + \frac{ 3\;
\sigma^{3}\sqrt{\pi}} {4\;r^{5}} \text{Erf}\left(\frac{r}{\sigma}\right) \right),
\ee
with $\delta_{jl}$ the Kronecker delta, $\bx=(x_1,x_2,x_3)^T$ and
$\text{Erf}(r)=\frac{2}{\sqrt{\pi}} \int_0^r  e^{-s^2} ds$
 the error function.
We choose $\sigma =1.4$ and compute the potential $\Phi(\bx)$ on a uniform mesh grid, i.e., $h_{x} = h_{y}=h_{z}$
on the domain $[-L,L]^{3}$ by the {\sl DST} and {\sl NUFFT} methods.
Table \ref{tab:diff:dip} shows the numerical errors $e_h$ via the {\sl DST} and {\sl NUFFT} methods
 with different dipole axis, i.e.,
$\bn=(0.82778,0.41505,-0.37751)^T$ and $\bm=(0.31180,0.93780,-0.15214)^T$, while
Table \ref{tab:same:dip} presents errors $e_h$ with the same dipole axis, i.e., $\bn=\bm=(0,0,1)^T$.

From Tabs. \ref{tab:diff:dip}-\ref{tab:same:dip},
we can clearly observe that:
(i) The errors are saturated in the {\sl DST} method as the mesh size $h$
tends smaller and the saturated accuracy decreases linearly with respect to the domain size $L$;
(ii) The NUFFT method is spectrally accurate and it essentially does not depend on the domain,
which implies that a very large bounded computational domain is not necessary in practical computations.

%%%%%%  Dipolar with non mean-zero Gaussian density: with different dipole axis %%%%
\begin{table}[h!]
\tabcolsep 0pt \caption{Errors of the 3D dipole interaction with different axis for different $h$ and $L$.}
\label{tab:diff:dip}
\begin{center}\vspace{-1em}
\def\temptablewidth{1\textwidth}
{\rule{\temptablewidth}{1pt}}
\begin{tabularx}{\temptablewidth}{@{\extracolsep{\fill}}p{1.35cm}rllll}
NUFFT & h=2&  $h = 1$  & $h= 1/2$ & $h=1/4$  \\\hline  %&   $h  = 1/8$
$L=4$   &	6.004E-01 & 6.122E-03	  &  	1.362E-04  &	9.823E-05	   \\ % &   8.599E-05
$L = 8$ & 6.344E-01 & 5.739E-03   &  1.189E-09   &   6.323E-14  \\ % &   6.293E-14
$L = 16$ & 6.641E-01  & 6.054E-03   &   1.162E-09   &   1.188E-13 \\[0.5em]%  &   1.189E-13
\hline
DST &  $h = 1$  & $h= 1/2$ & $h=1/4$ & $h  = 1/8$  \\\hline
$L= 8$ &1.985E-01 & 2.022E-01 & 2.038E-01 &2.046E-01\\
$L=16$ &7.135E-02 & 7.172E-02 & 7.200E-02 &7.214E-02\\
$L=32$ &2.622E-02 & 2.544E-02 & 2.549E-02 &2.552E-02\\
\end{tabularx}
{\rule{\temptablewidth}{1pt}}
\end{center}
\end{table}

%%%%%%  Dipolar with non mean-zero Gaussian density: the same dipole axis %%%%
\begin{table}[h!]
\tabcolsep 0pt \caption{Errors for the 3D dipole interaction with the same axis
for different $h$ and $L$.}
\label{tab:same:dip}
\begin{center}\vspace{-1em}
\def\temptablewidth{1\textwidth}
{\rule{\temptablewidth}{1pt}}
\begin{tabularx}{\temptablewidth}{@{\extracolsep{\fill}}p{1.35cm}rllll}
NUFFT & $h = 2$  & $h = 1$  & $h= 1/2$ & $h=1/4$ \\\hline    %&  $h  = 1/8$
$L=4$   & 1.118E-01 & 3.454E-04   &  1.335E-04	    &	1.029E-04	\\	%&  9.201E-05
$L = 8$ & 5.281E-02  & 3.428E-04     &  9.834E-12     &  1.601E-14    \\      %&   3.317E-14
$L = 16$ & 5.202E-02  & 3.551E-04      &  1.143E-11   &    8.089E-15     \\[0.5em]     %&   1E-15
\hline
DST &  $h = 1$  & $h= 1/2$ & $h=1/4$ & $h  = 1/8$  \\ \hline
$L= 8$ &6.919E-02 & 7.720E-02 & 8.124E-02 &8.327E-02\\
$L=16$ &2.709E-02 & 2.853E-02 & 2.925E-02 &2.961E-02\\
$L=32$ &1.008E-02 & 1.033E-02 & 1.046E-02 &1.052E-02\\
\end{tabularx}
{\rule{\temptablewidth}{1pt}}
\end{center}
\end{table}

\begin{exmp}\label{2d_exmp} {\sl Dipole-dipole interaction in 2D.} \end{exmp}
Here, we take $d=2$ and choose the source density as $\rho(\bx) = e^{-|\bx|^2/\sigma^2}$
with $\sigma>0$.  The exact 2D dipole interaction  with two  dipole orientations $\bn_\perp$ and $\bm_\perp$
can be  given as  \cite{BaoJiangLeslie}:
\bea
\Phi(\bx)&=&\frac{3\,\sqrt{\pi}\,e^{-r}}{4\sigma}\Big[(\bn_\perp\cdot \bm_\perp)(I_0(r)-I_1(r))
-\frac{2\,(\bx\cdot\bn_\perp)(\bx\cdot\bm_\perp)}{\sigma^2}\Big(I_0(r)\nn\\
&&-\frac{1+2\,r}{2\,r}I_1(r)\Big)\Big]
+ \frac{3\,\sqrt{\pi}\,n_3\,m_3\,r\,e^{-r}}{\sigma}\Big[I_0(r)-I_1(r) - \fl{I_0(r)}{2\,r}  \Big],
\eea
where  $r=\frac{|\bx|^2}{2\sigma^2}$,  $I_0$ and $I_1$ are the modified Bessel
functions of order $0$ and $1$, respectively \cite{handbook}.
Here, we choose $\sigma=\sqrt{1.3}$ and dipole axis as $\bn_\perp =(0, -0.896)^T$ and
$\bm_\perp=(0, -0.52476)^T$ (corresponding to $\bn =(0, -0.896, 0.44404)^T$ and
$\bm=(0, -0.52476, 0.85125)^T$ in 3D).
Table \ref{tab:2d_exmp} shows the errors $e_h$  via the {\sl DST} and {\sl NUFFT} methods
for different domain size $L$ and mesh size $h$.

From Tab. \ref{tab:2d_exmp},  we can clearly observe that:
(i) The errors are saturated in the {\sl DST} method as the mesh size $h$
tends smaller and the saturated accuracy decreases linearly with respect to the domain size $L$;
(ii) The NUFFT method is spectrally accurate and it essentially does not depend on the domain
if it is adequately large.

\begin{table}[h!]
\tabcolsep 0pt \caption{Errors of the 2D dipole interaction by different methods with $h$ on $[-L,L]^2$.}
\label{tab:2d_exmp}
\begin{center}\vspace{-1em}
\def\temptablewidth{1\textwidth}
{\rule{\temptablewidth}{1pt}}
\begin{tabularx}{\temptablewidth}{@{\extracolsep{\fill}}p{1.35cm}rllll}
NUFFT&  $h = 2$    & $h= 1$       & $h=1/2$     & $h  = 1/4$      \\\hline
$L=4$   &	 11.96   & 6.444E-01	& 5.251E-06	&  7.343E-06   \\
$L = 8$ &  1.279   & 1.611E-02   &  4.039E-07  & 4.720E-14   \\ % & 4.744E-14 \\%[0.5em]
$L = 16$  &3.289E-01  & 1.631E-02   & 4.226E-08   & 3.489E-14 \\[0.5em]%    & 3.516E-14
\hline
DST&  $h = 2$    & $h= 1$       & $h=1/2$     & $h  = 1/4$      \\ \hline
$L = 8$ & 3.200E-01 &  1.944E-02 &  1.145E-02&   1.208E-02 \\% &  1.239E-02  & 1.255E-02\\
$L = 16$& 3.264E-01 &  1.660E-02 &  2.971E-03&   3.048E-03 \\%&  3.087E-03  & 3.106E-03\\
$L = 32$& 3.281E-01 &  1.636E-02 &  7.590E-04&   7.686E-04 \\%&  7.734E-04  & 7.758E-04
\end{tabularx}
{\rule{\temptablewidth}{1pt}}
\end{center}
\end{table}

\begin{exmp} {\sl Dipole-dipole interaction for anisotropic densities.} \end{exmp}
In this example, we consider the dipole-dipole interaction for
anisotropic densities which are localized in
one or two spatial directions. As stated in the introduction,
the 3D/2D dipole interaction  \eqref{DDI-Eqv} can be
solved analytically via the Poisson equation \eqref{Laplace} and the fractional
Poisson equation \eqref{Sqare-Lap} in 3D and 2D, respectively.
Therefore, the dipole-dipole interaction can be obtained analytically via the solution of
the Poisson/ fractional Poisson potential, followed by differentiation.

{\sl \textbf{The 2D case}}. For an anisotropic density
$\rho(x,y) = \frac{1}{4\pi\varepsilon} e^{-\frac{x^2}{4} -\frac{y^2}{4\varepsilon^2}}$ with a small  parameter $0<\varepsilon\le 1$, the 2D Coulomb potential \eqref{Sqare-Lap} is given analytically
\cite{SP-NUFFT} as:
\bea\label{Coulomb:2D:aniso}
u(x,y) = \frac{1}{2\pi\sqrt{\pi}} \int_0^\infty  G(x,y,s)   d s, \quad \quad G(x,y,s)
=\frac{\text{exp}(-\frac{x^2}{4 (1+s^2)}-\frac{y^2}{4 (s^2+\varepsilon^2) }) }
{ \sqrt{s^2+1}\sqrt{s^2+\varepsilon^2}}.
\eea
Then, the 2D DDI can be obtained by differentiating
the integrand $G$ in \eqref{Coulomb:2D:aniso}.
For the convenience of readers, it can be evaluated explicitly as:
\bea\label{DDI:2D:aniso}
\Phi(x,y)=-\frac{3}{4\pi^{3/2}}\! \int_0^\infty  \left( ( n_1 m_1 -n_3 m_3 ) G_{xx} + ( n_2 m_2 -n_3 m_3 )  G_{yy} + (n_1 m_2 + n_2 m_1) G_{xy}\right)  ds.
\eea
Similarly as \cite{SP-NUFFT}, to numerically evaluate \eqref{DDI:2D:aniso}, we first split it into two integrals and reformulate the one with infinite interval into an equivalent integral with finite interval by a change of variable. Then, we apply high order Gauss-Kronrod quadrature to each integral to get reference solutions. Here we omit details for brevity. With this way, we can obtain the `exact' 2D DDI with the given density $\rho(x,y)$.

As discussed in \cite{SP-NUFFT}, the 2D Coulomb interaction can be well-resolved
by the NUFFT method on a heterogenous rectangle
$\mathcal{D}_\varepsilon = [-L,L] \times [-\varepsilon L , \varepsilon L ]$.
The DST method, best suited for solving the PDEs with homogeneous Dirichlet boundary condition on a rectangular domain, fails to produce even a satisfactory result on $\mathcal{D}_\varepsilon$, mainly because the potential does not decay fast enough.
Actually, the DST method can give reasonably accurate results on a
square $\mathcal{D} = [-L,L]^2$ due to that the homogeneous Dirichlet
boundary condition doesn't bring significant error, however, one needs to resolve the
anisotropic density with $h_y = \varepsilon h_x$.
Then the computational and storage costs for the DST method will
correspondingly scale linearly as a function of $1/\varepsilon$.
Here we adapt the similar strategy for the choice of computational domains for
the NUFFT and DST methods to compute the DDI.
Table \ref{tab:ani_2d} presents errors of the 2D dipole interaction for anisotropic densities by NUFFT on $\mathcal{D}_\varepsilon$ and DST on $\mathcal{D}$
with $h_x = 1/8, h_y =\varepsilon h_x $ for the same $\bn, \bm$ as in the Example \ref{2d_exmp}.

\begin{table}[h!]
\tabcolsep 0pt \caption{Errors of the 2D dipole interaction for anisotropic densities solved on $\mathcal{D}_\varepsilon$  and
$\mathcal{D}$ for the NUFFT and DST methods, respectively, with $h_x = 1/8$.}
\label{tab:ani_2d}
\begin{center}\vspace{-1em}
\def\temptablewidth{1\textwidth}
{\rule{\temptablewidth}{1pt}}
\begin{tabularx}{\temptablewidth}{@{\extracolsep{\fill}}p{1.35cm}rlllll}
NUFFT	&  $\varepsilon = 1$ & $\varepsilon = 1/2$ &$\varepsilon = 1/4$ &$\varepsilon = 1/8$ &$\varepsilon = 1/16$ \\ \hline
$L = 8$ & 3.456E-07& 4.167E-07 & 3.984E-07 &3.207E-07& 2.864E-07 \\
$L = 16$& 1.005E-12& 7.531E-15 & 5.119E-15 &4.108E-15& 3.720E-15 \\
$L = 32$& 2.241E-12& 6.856E-15 & 4.913E-15 &4.072E-15& 3.855E-15 \\[0.5em]
\hline
DST&  $\varepsilon = 1$ & $\varepsilon = 1/2$ &$\varepsilon = 1/4$ &$\varepsilon = 1/8$ &$\varepsilon = 1/16$ \\  \hline
$L = 8$ &4.014E-02& 3.182E-02& 1.711E-02&  6.276E-03 &2.181E-03\\
$L = 16$&9.604E-03& 7.534E-03& 4.035E-03&  1.479E-03 &5.137E-04\\
$L = 32$&2.386E-03& 1.868E-03& 9.995E-04&  3.661E-04 &1.272E-04\\
\end{tabularx}
{\rule{\temptablewidth}{1pt}}
\end{center}
\end{table}

\vspace{0.5cm}
As for {\sl \textbf{the 3D case}}, there are two typical kinds of anisotropic densities, that is, densities that are strongly localized in one or two directions.
The first typical kind of anisotropic density is localized in one direction. For example, choose the density as  $\rho(x,y,z) = \frac{1}{8\pi\sqrt{\pi}\varepsilon} e^{-\frac{x^2+y^2}{4}}e^{-\frac{z^2}{4\varepsilon^2}}$, and  its corresponding 3D Coulomb potential \eqref{Laplace} is given as:
\bea\label{Coulomb:3D:aniso:z}
u(x,y,z) = \frac{1}{8\pi^{3/2}}\int_0^\infty e^{-\frac{x^2+y^2}{4(1+s)}} e^{-\frac{z^2}
{4(s+\varepsilon^2)}} \frac{1}{ (1+s) \sqrt{s+ \varepsilon^2}} d s.
\eea
The second kind of anisotropic density is localized in two directions.
For example, the density is taken as $\rho(x,y,z) = \frac{1}{8\pi\sqrt{\pi}\varepsilon^2} e^{-\frac{x^2+y^2}{4\varepsilon^2}}e^{-\frac{z^2}{4}}$, and the corresponding
3D Coulomb potential \eqref{Laplace}  is given
analytically as:
\bea\label{Coulomb:3D:aniso:xy}
u(x,y,z) = \frac{1}{8\pi^{3/2}}\int_0^\infty e^{-\frac{x^2+y^2}{4(s+\varepsilon^2)}}
e^{-\frac{z^2}{4(s+1)}} \frac{1}{ \sqrt{1+s}\;(s+\varepsilon^2)}ds.
\eea
Similarly as in the 2D case, the DDI in 3D can be obtained by
differentiating the integrand in \eqref{Coulomb:3D:aniso:z} and \eqref{Coulomb:3D:aniso:xy}.
They can be evaluated numerically in a similar way, which can be viewed as the `exact' solution.
For brevity, we omit the formulas and relevant details.

To numerically compute the 3D dipole interaction by the NUFFT and DST methods,
we shall adopt the meshing strategy, i.e., $h_y = h_x$ and $h_z = \varepsilon h_x$ for densities localized only in $z$-direction and $ h_y = h_x =  \varepsilon h_z$ for densities localized  in $x,y$ directions. Similarly, the NUFFT method is applied on a heterogenous cube $\mathcal{D}_\varepsilon = [-L,L]^2 \times [-\varepsilon L,\varepsilon L]$ or $[-\varepsilon L,\varepsilon L]^2 \times [- L,L]$.
The DST method is used on $\mathcal{D}= [-L,L]^3$ so that the homogeneous Dirichlet
boundary condition doesn't bring significant error.
Correspondingly, the computational and storage costs of the
DST method will  scale linearly as a function
of $1/\varepsilon$ (the first kind) or $1/\varepsilon^2$ (the second kind).

To show the accuracy performance of both methods, we take the first kind density as the test function \eqref{Coulomb:3D:aniso:z}.  Here
we take the same dipole axis $\bn = \bm = (0,0,1)^T$ for simplicity.  Table \ref{tab:ani_3d_cs1} presents errors of the 3D dipole interaction for anisotropic densities localized in the $z$-direction by NUFFT on $\mathcal{D}_\varepsilon$ and DST on $\mathcal{D}$
with $h_x = 1/4$ and $h_z =\varepsilon h_x $.

\begin{table}[h!]
\tabcolsep 0pt \caption{Errors of the 3D dipole interaction  \eqref{Coulomb:3D:aniso:z} by NUFFT  on $\mathcal{D}_\varepsilon$ and DST on
$\mathcal{D}$ with $h_x = 1/4$ and $\bn = \bm = (0,0,1)^T$.}
\label{tab:ani_3d_cs1}
\begin{center}\vspace{-1em}
\def\temptablewidth{1\textwidth}
{\rule{\temptablewidth}{1pt}}
\begin{tabularx}{\temptablewidth}{@{\extracolsep{\fill}}p{1.35cm}rlllll}
NUFFT	&  $\varepsilon = 1$ & $\varepsilon = 1/2$ &$\varepsilon = 1/4$ &$\varepsilon = 1/8$ &$\varepsilon = 1/16$ \\ \hline
$L = 8$   &3.004E-08 & 2.581E-08 &2.307E-08 & 1.988E-08 &1.578E-08\\
$L = 16$  &1.598E-14 & 7.590E-15 &4.590E-15 & 2.184E-15 &1.193E-15   \\ [0.5em]
\hline
DST&  $\varepsilon = 1$ & $\varepsilon = 1/2$ &$\varepsilon = 1/4$ &$\varepsilon = 1/8$ &$\varepsilon = 1/16$ \\  \hline
$L = 8$  & 1.409E-01 & 7.667E-02 &4.308E-02 &2.633E-02 &1.716E-02\\
$L = 16$ & 5.003E-02 & 2.754E-02 &1.548E-02 &9.453E-03 &6.159E-03\\
$L = 32$ & 1.786E-02 & 9.836E-03 &5.522E-03 &3.370E-03 &2.195E-03\\
\end{tabularx}
{\rule{\temptablewidth}{1pt}}
\end{center}
\end{table}

From Tabs. \ref{tab:ani_2d}-\ref{tab:ani_3d_cs1}, we can conclude: (1) the NUFFT
can evaluate accurately the 2D and 3D dipole  interaction with anisotropic densities.
(2) The DST method can still capture satisfactory results if the computational domain
is large enough, however, the computational and storage costs increase
when the heterogeneity of the density increases, which makes it less applicable, especially in 3D simulation.

%%%%%%%%%%%%%%%%%%%%%%%%%%%%%%%%%%%%%%%%%%%%%%
\section{Ground state computation }
In this section, we propose an efficient and accurate numerical method for computing the ground state
by combining  the normalized gradient flow which is discretized  by the
backward Euler Fourier pseudospectral method
and the NUFFT nonlocal DDI interaction solver. We shall refer to this new method as {\sl GF-NUFFT} hereafter.
The spatial spectral accuracy is investigated in details, the
{\sl virial} identity is verified numerically,  with comparison to some existing results in
\cite{DipJCP}, to show the advantage of the
{\sl GF-NUFFT} method in term of accuracy.

\subsection{A numerical method via the NUFFT}
Let $\Delta t>0$ be the time step and denote $t_n=n\Delta t$ for $n=0,1,2,\ldots$\,\,.
Many efficient and accurate numerical methods have
been proposed for computing the ground state of the GPE \cite{DipJCP,BCL_JCP,BD_SISC,SPMCompare}.
One of the most simple and successful method is by solving the following
 gradient flow with discretized normalization (GFDN):
\bea \label{gf-eq1}
&&\partial_t \phi(\bx,t)= \left[\frac{1}{2} \nabla^2  -V(\bx)- \beta |\psi|^2 - \lambda \,\Phi(\bx,t) \right]
\phi(\bx,t),\quad \bx \in {\mathbb{ R}}^d,\quad t_n\le t<t_{n+1},\\
\label{gf-colb1}
&& \Phi(\bx,t) = \left(U_{\rm dip}\ast |\phi|^2\right)(\bx,t), \;\;\;\quad\quad \qquad \qquad \qquad \qquad
\bx \in {\mathbb{ R}}^d, \quad t_n\le t< t_{n+1},\\
\label{gffg3}
&&\phi(\bx,t_{n+1})=\frac{\phi(\bx,t_{n+1}^-)}{\|\phi(\bx,t_{n+1}^-)\|}, \qquad \quad \qquad \qquad \qquad \qquad
\bx \in {\mathbb{ R}}^d, \quad n\ge0,
\eea
with the initial data
\be
\phi(\bx,0)=\phi_0(\bx), \qquad \bx \in {\mathbb{ R}}^d, \qquad {\rm with} \qquad \|\phi_0\|^2:=\int_{{\mathbb R}^d}
|\phi_0(\bx)|^2\,d\bx=1.
\ee
Let $\phi^n(\bx)$ and $\Phi^n(\bx)$ be the numerical approximations of $\phi(\bx,t_n)$ and
$\Phi(\bx,t_n)$, respectively, for $n\ge0$.
The above GFDN is usually discretized in time via the backward Euler method \cite{DipJCP,SPMCompare}
\bea \label{gf-eq2}
&&\frac{\phi^{(1)}(\bx)-\phi^{n}(\bx)}{\Delta t}= \left[\frac{1}{2} \nabla^2
-V(\bx)- \beta |\psi^n|^2-\lambda \,\Phi^n(\bx) \right]
\phi^{(1)}(\bx),\qquad \bx \in {\mathbb{ R}}^d,\\
\label{gf-colb2}
&& \Phi^n(\bx) = \left(U_{\rm dip}\ast |\phi^n|^2\right)(\bx),
\quad\qquad\qquad\qquad \qquad \quad \qquad \qquad\;\;\,
\bx \in {\mathbb{ R}}^d, \\
\label{gffg33}
&&\phi^{n+1}(\bx)=\frac{\phi^{(1)}(\bx)}{\|\phi^{(1)}(\bx)\|}, \quad \qquad\qquad \qquad \qquad \qquad \qquad
\bx \in {\mathbb{ R}}^d, \quad n\ge0.
\eea
As it is known, the ground state  decays exponentially fast due to the trapping potential,
therefore, in practical computations,
we shall first truncate the whole space to a bounded domain $\mathcal{D}$  and impose periodic boundary conditions.
It is worthwhile to point out that the dipole interaction is originally defined by convolution, therefore it does not require any boundary condition.
Then, the equation (\ref{gf-eq2}) is discretized in space via the Fourier pseudospectral method and the dipole interaction $\Phi^n(\bx)$ in (\ref{gf-colb2})
is evaluated by the NUFFT solver.  The initial guess $\phi_0(\bx)$ is usually chosen as a positive function, e.g., a Gaussian, and the ground state $\phi_g(\bx)$
is obtained numerically as $\phi^n(\bx)$ once $\frac{\| \phi^n(\bx)-\phi^{n+1}(\bx)\|_{l^\infty}}{\Delta t}
\le \varepsilon_0 $ is satisfied, where $\varepsilon_0$ is the desired accuracy.
The details are omitted here for brevity.

\subsection{Numerical results }
%\textcolor{red}
%{Should we also take the comparison with GF-DST like what have been done in the Coulomb case?}}

In order to study the spatial accuracy of the {\sl GF-NUFFT} method for computing the ground state,
we denote $\Phi_g(\bx)=(U_{\rm dip}\ast|\phi_g|^2)(\bx)$ and introduce the error functions
\[e_{\phi_g}^h:=\frac{\|\phi_g(\bx)-\phi_g^h(\bx)\|_{l^2}}{\|\phi_g(\bx)\|_{l^2}} ,
\qquad e_{\Phi_g}^h:=\frac{\|\Phi_g(\bx)-\Phi_g^h(\bx)\|_{l^2}}{\|\Phi_g(\bx)\|_{l^2}},
\]
where $\phi_g^h$ and $\Phi_g^h$ are obtained numerically by a numerical method
with the mesh size $h$.
%very fine mesh size, i.e., $h = 1/16$ in this section.
Additionally, we split the energy functional into four parts
\[E(\phi)=E_{\rm kin}(\phi)+E_{\rm pot}(\phi)+E_{\rm int}(\phi)+E_{\rm dip}(\phi),\]
where the   kinetic energy  $E_{\rm kin}(\phi)$, the potential energy  $E_{\rm pot}(\phi)$, the interaction
energy $E_{\rm int}(\phi)$, and the dipole interaction energy $E_{\rm dip}(\phi)$ are defined as
\beas
& E_{\rm kin}(\phi)   =  \frac{1}{2} \int_{\mathbb R^d} |\nabla \phi(\bx)|^2 d \bx,\quad
E_{\rm pot}(\phi)  =  \int_{\mathbb R^d} V(\bx) |\phi(\bx)|^2 d \bx, \\
& E_{\rm int}(\phi)  =  \frac{\beta}{2}\int_{\mathbb R^d} |\phi(\bx)|^4  d\bx, \qquad
E_{\rm dip}(\phi)  =  \frac{\lambda}{2}\int_{\mathbb R^d}\Phi(\bx) |\phi(\bx)|^2  d\bx,
\eeas
respectively. Moreover, the chemical potential can be reformulated as
$\mu(\phi)=E(\phi)+E_{\rm int}(\phi)+E_{\rm dip}(\phi)$. Furthermore, if
 the external potential $V(\bx)$ in (\ref{DipGPE0}) is taken as the harmonic
 potential \eqref{Vpoten} \cite{BC2013,BJNY,SP-NUFFT,ExtBd}, the energies of the ground state satisfy  the following {\sl virial} identity
\be\label{vir_iden}
0=I:=2 E_{\rm kin}(\phi_g)  - 2 E_{\rm pot}(\phi_g)  + 3 E_{\rm int}(\phi_g) +3 E_{\rm dip}(\phi_g).
\ee
We denote $I^h$ as an approximation of $I$ when $\phi_g$ and $\Phi_g$ are
replaced by $\phi_g^h$ and $\Phi_g^h$ in (\ref{vir_iden}).
In our computations, the ground state $\phi_g^h$ is reached numerically when
$\frac{\Vert \phi^{n+1}(\bx)-\phi^n(\bx)\Vert_{l^\infty}}{\Delta t}\le \varepsilon_0$ with
 $\varepsilon_0 =10^{-10}$.  The initial data $\phi_0(\bx)$ is chosen as a Gaussian and the
 time step is taken as $\Delta t=10^{-2}$.  In the comparisons, the  ``exact" solution $\phi_g(\bx)$ was
obtained numerically via the GF-NUFFT method on a  large enough domain $\Omega$ with
 small enough mesh size $h$ and the same time step $\Delta t =10^{-2}$.

\begin{table}[h!]
\tabcolsep 0pt \caption{Errors of the ground states and the dipole interaction obtained
by the GF-NUFFT method for $\bn = (0,0,1)^T$ and $\beta = 200$ with different mesh sizes $h$ and
$\lambda$.   }
\label{tab:gs:ddi:acc}
\begin{center}\vspace{-1em}
\def\temptablewidth{1\textwidth}
{\rule{\temptablewidth}{1pt}}
\begin{tabularx}{\temptablewidth}{@{\extracolsep{\fill}}p{0.025cm}llllll}
\multicolumn{2}{c|}{GF-NUFFT} & $h=2$ &  $h = 1$  & $h =1/2$ & $h=1/4$  & $h=1/8$ \\ \hline
\multirow{3}*{\scalebox{1.25}{$e^h_{\phi_g}$}}
&\multicolumn{1}{|l|}{\;\;$\lambda=-100\;$}  & 1.783E-02&   3.102E-03  & 3.463E-05 &  3.652E-09 &  4.133E-12 \\
&\multicolumn{1}{|l|}{\;\;$\lambda=\;\;100\;$}& 1.263E-02&   2.717E-03  & 3.599E-05 &  6.183E-09 &  2.841E-12\\
&\multicolumn{1}{|l|}{\;\;$\lambda=\;\;200\;$}& 1.670E-02&   3.049E-03  & 8.897E-05 &  5.364E-08 &  3.871E-12\\ \hline
\multirow{3}*{\scalebox{1.25}{$e^h_{\Phi_g}$}}
&\multicolumn{1}{|l|}{\;\;$\lambda=-100\;$}  & 2.810E-02&   3.683E-03  & 1.842E-05 &  1.555E-09 &  8.132E-12  \\
&\multicolumn{1}{|l|}{\;\;$\lambda=\;\;100\;$}& 2.385E-02&   4.932E-03  & 9.445E-05 &  1.996E-08 &  2.750E-12 \\
&\multicolumn{1}{|l|}{\;\;$\lambda=\;\;200\;$}& 1.406E-02&   5.681E-03  & 2.424E-04 &  1.921E-07 &  3.121E-12\\
\end{tabularx}
{\rule{\temptablewidth}{1pt}}
\end{center}
\end{table}

\textbf{Accuracy confirmation}. To show the accuracy of the GF-NUFFT, we take $d=3$,
$\bn = (0,0,1)^T$, $\beta = 200$ and $V(\bx) = \frac{1}{2}(x^2+y^2+z^2/4)$.
Table \ref{tab:gs:ddi:acc} presents errors of
the ground states and the corresponding dipole interactions computed
on a fixed domain $[-8,8]^3$ with different mesh sizes and $\lambda$.
From this Table,  we can observe clearly the spectral convergence in space of the GF-NUFFT method.

\textbf{Virial identity}. Here we take the same physical parameters as used in \cite{DipJCP} (cf. Table 3), i.e.,
$d=3$, $\beta = 207.16$ and $V(\bx) = \frac{1}{2}(x^2+y^2+z^2/4)$.
We compute the ground state on a larger domain, i.e., $[-12,12]^3$, with a coarser mesh size $h_x=h_y=h_z = 1/4$.
Different energies of the ground state and related quantities are shown in Table \ref{virial_dip3d}.
Compared with Table 3 in \cite{DipJCP} where the identity is only accurate up to 3 significant digits, our results by
the GF-NUFFT method agree quite well with the identity, up to 9 significant digits.

%%%%  Virial identity compared with Table 3 in Wang Cai Bao's JCP paper
%
% Computational details : [-12,12]^3 with h_x = h_y = h_z = 1/4 on grid 96^3, absolute error control is 1e-12.
%
%
\begin{table}[htbp]
\tabcolsep 0pt \caption{Different energies of the ground state and $I^h$ for
the 3D dipolar BEC with $\beta =  207.16$ for different $\lambda$.}
\label{virial_dip3d}
\begin{center}\vspace{-0.5em}
\def\temptablewidth{1\textwidth}
{\rule{\temptablewidth}{1pt}}
\begin{tabularx}{\temptablewidth}{@{\extracolsep{\fill}}p{1.25cm}lllllll}
$\lambda$  &$E_g$ &$\mu_g$ &$E_{\textrm {kin}}^g$  & $E_{\textrm {pot}}^g$ & $E_{\textrm {int}}^g$ & $E_{\textrm {dip}}^g$&$ I^h $ \\
\hline
$-103.58 $ &2.9584  & 3.9301  & 0.26466 &  1.7221  & 0.83892 &  0.13273  & 6.6214E-10  \\
$-51.79$ &2.8841  & 3.8187  & 0.27379 &  1.6757  & 0.85255 &  0.082056  & 5.7861E-10  \\
$0$     &2.7943  & 3.6830  & 0.28621 &  1.6193  & 0.88875 &  0.0000       & 5.0929E-10  \\
$51.79$  &2.6875  & 3.5201  & 0.30303 &  1.5519  & 0.94903 & -0.11646  & 4.5134E-10  \\
$103.58$  &2.5593  & 3.3213  & 0.32704 &  1.4701  & 1.0451     & -0.28304  & 3.6672E-10  \\
$155.37$  &2.3998  & 3.0674  & 0.36538 &  1.3668  & 1.2105     & -0.54290  & 2.3288E-10  \\
$207.16$     &2.1838  & 2.7011  & 0.44525 &  1.2212  & 1.5749     & -1.0576      &-1.6697E-10  \\
\end{tabularx}
{\rule{\temptablewidth}{1pt}}
\end{center}
\end{table}

%%%%%%%%%%%%%%%%%%%%%%%%%%%%%%%%%%%%%%%%%%%%%%
%%%%%%%%%%%%%%%%%%%%%%%%%%%%%%%%%%%%%%%%%%%%%%
\section{Dynamics simulation}
In this section,
instead of solving the GPE (\ref{DipGPE0})-(\ref{ini-con0}), we consider a more general GPE
in $d$-dimensions ($d=2,3$) with both the damping term and time (in-)dependent DDI:
\bea\label{DipGPE1}
&& i\p_t \psi({\bx}, t) = \left[-\fl{1}{2}\nabla^2 + V({\bf x}) + \beta |\psi|^{2\sigma} +
\lambda\Phi(\bx,t)- if(|\psi|^2) \right]\psi(\bx,t),\\
\label{dipole-poten1}
&&\Phi(\bx,t)=(U_{\rm dip}\ast |\psi|^2)(\bx,t), \qquad \bx\in {\mathbb R}^d, \qquad t\ge0,\\
\label{ini-new}
&& \psi(\bx,t=0)=\psi_0(\bx).
\eea
Here,   $\sigma >0$ corresponds to
the type of the nonlinearity ($\sigma =1$ represents to the cubic nonlinearity, and resp.,
$\sigma=2$ to a quintic nonlinearity).   $f(\rho)\ge 0$ for $\rho=|\psi|^2\ge 0$ is a real-valued
monotonically increasing function that represents the type of damping.  In BEC,
when $f(\rho)\equiv 0$, (\ref{DipGPE1}) reduces to the usual GPE (\ref{DipGPE0}) without damping effect, while
a  linear damping term $f(\rho)\equiv\delta$
with $\delta>0$ represents inelastic collisions with the background gas.  In addition,  the cubic damping
$f(\rho)=\delta_1 \rho$ with $\delta_1>0$ describes two-body loss,  a quintic
damping term of the form $f(\rho)=\delta_2 \rho^2$
with $\delta_2>0$ corresponds to the three-body loss, and their combination $f(\rho)=\delta_1 \rho+\delta_2 \rho^2$
takes both the two and three-body loss into account.
Furthermore, the kernel of the dipole interaction may be time (in-)dependent, which is defined as
\bea\label{dipole-time}
U_{\rm dip}(\bx, t)&=& \fl{3}{4\pi}\fl{\bm(t)\cdot\bn(t)-3(\bx\cdot\bm(t))(\bx\cdot\bn(t))/|\bx|^2}{|\bx|^3}\\\nn
&=&-(\bm(t)\cdot\bn(t))\delta(\bx)-3\partial_{\bm(t)\bn(t)}\left(\fl{1}{4\pi|\bx|}\right),
\qquad  \bx\in{\Bbb R}^3,
\eea
where  $\bm(t) = (m_1(t), m_2(t), m_3(t))^T$ and $\bn(t) = (n_1(t), n_2(t), n_3(t))^T\in{\Bbb R}^3$
are two given time (in-)dependent unit vectors, representing  the  two dipole orientations.
The energy is modified as:
\bea
\nn
\mathcal{E}(t)
&=:&\int_{{\Bbb R}^d}\Big[ \fl{1}{2}|\nabla\psi|^2 + V({\bf x})|\psi|^2+\fl{\bt}{\delta+1}
|\psi|^{2\,(\delta+1)}+\fl{\lambda}{2}\Phi(\bx,t)\, |\psi|^2 \\[0.5em]
&&-\fl{\lambda}{2}\int_{0}^{t} \Big(\partial_s
U_{\rm dip}\ast |\psi|^2\Big) |\psi(\bx,s)|^2ds   \Big]d{\bf x},
\eea
which satisfies the following dynamical law:
\be\label{Energy_Disp}
\fl{d}{d t}\mathcal{E}(t) = -2\int_{{\Bbb R}^d} f(|\psi|^2) {\rm Im} \big( \psi\p_t\overline{\psi}\big) d\bx.
\ee
where $\overline{\psi}$ denotes the complex conjugate of $\psi$.
The total mass $N(t)$ (\ref{norm}) is dissipated as:
\be\label{Mass_Disp}
 \fl{d}{d t}N(t) =-2\int_{{\Bbb R}^d} f(|\psi|^2) |\psi|^2 d\bx.
 \ee

  We will present an accurate and efficient numerical method for simulating the
dynamics of the GPE  (\ref{DipGPE1})-(\ref{ini-new}). The method incorporates the NUFFT
solver for the evaluation of the nonlocal dipole interaction and the time-splitting Fourier pseudospectral
discretization for the GPE (\ref{DipGPE1}).

\subsection{Numerical method}

In practical computation,
we first truncate the problem (\ref{DipGPE1})-(\ref{ini-new}) into
a bounded computational domain $ \mathcal{D} = [L_x, R_x]\times[L_y, R_y]\times[L_z, R_z]$
if $d=3$,  or $ \mathcal{D} = [L_x, R_x]\times[L_y, R_y]$ if $d=2$. From $t=t_n$
to $t=t_{n+1}$, the GPE (\ref{DipGPE1}) will be solved in two steps. One first solves
\be\label{1step}
i\p_t\psi(\bx, t) = -\fl{1}{2}\nabla^2\psi(\bx,t), \qquad  \quad
\bx\in{\mathcal D}, \qquad t_n\leq t\leq t_{n+1},
\ee
with periodic boundary condition  on the boundary $\p\mathcal{D}$
for a time step of length $\Dt t$, followed by solving
\bea
\label{2stepA}
 i\p_t \psi({\bx}, t)&=&\left[ V({\bf x}) + \beta |\psi|^{2\sigma} +
\lambda\Phi(\bx,t)- if(|\psi|^2) \right]\psi(\bx,t),  \qquad \bx\in \mathcal{D},
\quad t_n\leq t\leq t_{n+1}, \qquad  \\
\label{2stepB}
\Phi(\bx,t)&=&\big(U_{\rm dip}\ast |\psi|^2\big) (\bx, t), \qquad\qquad\quad
 \bx\in \mathcal{D}, \quad t_n\leq t\leq t_{n+1},
\eea
for the same time step.
The linear subproblem (\ref{1step}) will be discretized in space by the Fourier pseudospectral
method  and integrated in time exactly in the phase space, while the nonlinear subproblem
(\ref{2stepA})-(\ref{2stepB})  can be integrated exactly,
one can refer to  \cite{DipJCP, BMTZ2013, BJ2003, BJM2004} for details.
To simplify the presentation, we will only present the scheme for the 3D case.
As for the 2D case, one can modify the algorithm straightforward.

 Let $L$, $M$, $N$ be even positive integers, choose $h_x=\fl{R_x-L_x}{L}$, $h_y=\fl{R_y-L_y}{M}$ and $h_z=\fl{R_z-L_z}{N}$
 as the spatial mesh sizes in $x$-, $y$-, and $z$- directions, respectively.  Define the index and grid points sets as
\beas
{\mathcal T}_{LMN} &=& \left\{(l, k, m)\,|\,0\leq l\leq L, \ 0\leq k\leq M, \ 0\leq m\leq N\right\}, \\
\widetilde{\mathcal T}_{LMN} &=& \left\{(p, q, r)\,|\,-\frac{L}{2}\leq p\leq \frac{L}{2}-1,
\ -\frac{M}{2}\leq q\leq \frac{M}{2}-1, \ -\frac{N}{2}\leq r\leq \frac{N}{2}-1\right\},\\
{\mathcal G}_{xyz} &=& \left\{ (x_l, y_k, z_m) =: (L_x + j h_x, L_y + k h_y, L_z + m h_z ),\ (l, k, m) \in  {\mathcal T}_{LMN}  \right\}.
\eeas
Define the functions
\[W_{pqr}^s(x,y,z)=e^{i \mu_p^x(x-L_x)}\,e^{i \mu_q^y(y-L_y)}\,e^{i \mu_r^z(z-L_z)},
\quad (p, q,r)\in\widetilde{\mathcal T}_{LMN},
\]
with
\[  \mu_p^x = \fl{2p\pi}{R_x-L_x}, \;\; \mu_q^y = \fl{2q\pi}{R_y-L_y},\;\;
\mu_r^z = \fl{2r\pi}{R_z-L_z},   \quad (p, q, r)\in \widetilde{\mathcal T}_{LMN}.\]
Let $\psi_{lkm}^n$ be the approximation of $\psi(x_l, y_k, z_m, t_n)$ for
$ (l, k,m) \in {\mathcal T}_{LMN}$ and $n\ge0$ and denote $\psi^n$ be the solution vector
at time $t=t_n$ with components $\{\psi_{lkm}^n, \ (l,k,m)\in {\mathcal T}_{LMN}\}$.
Taking the initial data as
$\psi_{lkm}^0=\psi_0(x_l, y_k, z_m)$ for $(l,k,m)\in {\mathcal T}_{LMN}$,
for $n\ge0$,  a second-order time splitting Fourier pseudospectral (TSFP) method to solve the GPE (\ref{DipGPE1})-(\ref{ini-new})
reads as:
\bea
\psi_{lkm}^{(1)}&=&\sum_{p=-L/2}^{L/2-1}\sum_{q=-M/2}^{M/2-1}\sum_{r=-N/2}^{N/2-1}
e^{-\fl{i\Dt t}{4} \left[ (\mu_p^x)^2+ (\mu_q^y)^2+(\mu_r^z)^2\right]}
\widehat{(\psi^n)}_{pqr}\; W_{pqr}^s(x_l,y_k,z_m), \nonumber\\[0.5em]
\psi_{lkm}^{(2)}&=&\psi_{lkm}^{(1)}\exp\left\{ -i\left[ \Dt t V(\bx) +  H(|\psi_{lkm}^{(1)}|^2, \Dt t) +
G(|\psi^{(1)}|^2, t^n, t^{n+1})(x_l,y_k,z_m)  \right] \right\}\nonumber\\
&&\qquad \qquad  \times \exp\{ -F(|\psi_{lkm}^{(1)}|^2, \Dt t) \}, \qquad (l,k,m)\in {\mathcal T}_{LMN}, \nonumber \\[0.5em]
\psi_{lkm}^{n+1}&=&\sum_{p=-L/2}^{L/2-1}\sum_{q=-M/2}^{M/2-1}\sum_{r=-N/2}^{N/2-1}e^{-\fl{i\Dt t}{4} \left[ (\mu_p^x)^2+ (\mu_q^y)^2+(\mu_r^z)^2\right]}
\widehat{(\psi^{(2)})}_{pqr}\; W_{pqr}^s(x_l,y_k,z_m).
\label{tssp2}
\eea
Here, $\widehat{(\psi^n)}_{pqr}$ and $ \widehat{(\psi^{(2)})}_{pqr}$ are
the discrete Fourier transform coefficients of
the vectors $\psi^n$ and $\psi^{(2)}$,
respectively, and the  functions  $H(\varphi,s)$, $G(\varphi, s, s_1)$ and $F(\varphi, s)$  are defined as:
\bea
\label{fun1}
&& H(\varphi, s)=\beta \int_{0}^{s} \left[h(\varphi,\tau)\right]^{\sigma} d\tau,   \qquad  F(\varphi, s) = \int_{0}^{s} f(h(\varphi, \tau)) d\tau,\\
\label{fun2}
&& G(\varphi, s, s_1)(\bx)=\lambda \int_{s}^{s_1} \left(U_{\rm dip}(\cdot, \tau)
\ast h(\varphi(\cdot,\tau - s)\right)(\bx) d\tau,
\eea
with
 \be
 h(\varphi,s)=\left\{ \begin{array}{cc}
 g^{-1}(g(\varphi)-2 s),   &   \varphi>0, \quad s \ge 0,    \\
  0,				   &     \varphi=0, \quad s\ge0,
 \end{array}  \right.
   \qquad
   g(s)=\int\fl{1}{s f(s)}.
 \ee

For a given damping function $f(s)$, in general,
 $g^{-1}(s)$ and thus $h(\varphi,s)$ may not have explicit expressions.
 In practical  computation,  one could
solve $h(\varphi, s)$ numerically from an auxiliary  ODE, and then evaluate (\ref{fun1})-(\ref{fun2}) via a numerical quadrature.
For details, one can refer to {\it Remark 2.1} in \cite{BJ2003}.  However, if the dipole axis is time independent, i.e.,
 $U_{\rm dip}(\bx, t)\equiv U_{\rm dip} (\bx,t=0) =: U_{\rm dip}^0(\bx)$,
for those damping terms that are frequently used in the physics
literatures, the
functions $H$, $F$ and $G$ can be integrated analytically.
For the convince of the reader, we list them here briefly \cite{BJ2003}:
 \begin{itemize}
\item Case I.  $f(\rho)\equiv 0$, i.e., no damping term, we have
\[H(\varphi, s)=\beta \varphi^\sigma s,  \qquad   F(\varphi, s) = 0, \qquad
G(\varphi, s, s_1)(\bx)=\lambda (s_1-s) \, \left(U_{\rm dip}^0 \ast \varphi\right)(\bx).
\]

\item Case II.  $f(\rho)=\delta >0$, i.e., the linear damping,  we have
\beas
&&H(\varphi, s) = \fl{\beta \varphi^\sigma }{2 \delta \sigma} \left(1-e^{-2 \delta \sigma s}\right),  \qquad F(\varphi, s)=\delta s, \\
&&G(\varphi, s, s_1)(\bx)= \fl{\lambda }{2 \delta}\left(1- e^{-2 \delta(s_1-s)}\right)\left(U_{\rm dip}^0 \ast \varphi\right)(\bx).
\eeas

\item Case III. $f(\rho)=\delta \rho^{q}$ with $\delta, q>0$, which corresponds to two ($q=1$) or three ($q=2$) body loss of particles, we have
\bea
\nn
&&F(\varphi, s)=\fl{1}{2q} \ln (1+2q\delta s \varphi^q ), \\[0.4em]
\nn
&&H(\varphi, s)= \left\{
\begin{array}{ll}
\fl{\beta}{2q\delta}\ln(1+2q\delta s \varphi^q ),  & \quad {\rm if} \quad \sigma=q, \\[0.4em]
\fl{\beta \varphi^{\sigma-q}}{2\delta(q-\sigma)}\left[\left(1+2q\delta s \varphi^q \right)^{1-\sigma/q}-1\right], &\quad  {\rm if}\quad \sigma\ne q,
\end{array} \right.   \\[0.4em]
\nn
&& G(\varphi, s, s_1)(\bx)= \lambda\, U_{\rm dip}^0 \ast \left\{
\begin{array}{ll}
\fl{1}{2\delta}\ln(1+2\delta(s_1-s) \varphi   ),  & \quad {\rm if} \quad  q=1, \\[0.4em]
\fl{\left(1+2q\delta (s_1-s) \varphi^q \right)^{1-1/q}-1}{2\delta(q-1) \varphi^{q-1}}, &\quad  {\rm if}\quad q \ne 1.
\end{array} \right.
\eea
\end{itemize}
The function $G$ is evaluated by the algorithm via the  NUFFT as discussed in previous sections,
and this method for discretizing the GPE (\ref{DipGPE1})-(\ref{ini-new}) is referred as {\sl TS-NUFFT}.

 \subsection{Test of the accuracy}
In this section, we first test the accuracy of our numerical method for computing the dynamics of
 the dipolar BEC.
To demonstrate the results,  we define the following error function
\be
e_{\psi}^{h,\, \Dt t}(t_n):=\frac{\Vert\psi(\bx,t_n)-\psi^n_{h,\, \Dt t}(\bx)\Vert_{l^2}}{\Vert\psi(\bx,t)\Vert_{l^2}},
\qquad n\ge0,
\ee
where $\Vert\cdot\Vert_{l^2}$ represents the $l^2$ norm, $\psi^n_{h,\, \Dt t}(\bx)$ is the numerical
approximation of $\psi(\bx, t=t_n)$  obtained by the {\sl TS-NUFFT} method (\ref{tssp2}) with mesh size
$h$ and time step $\Dt t$.   In this subsection, all examples are carried out for dipolar BEC without
damping effect, i.e., $f(\rho)\equiv 0$ in the GPE (\ref{DipGPE1}). Moreover, the computational domain
$\mathcal{D}$, the trapping potential $V(\bx)$ and the initial data $\psi_0(\bx)$ are respectively
chosen  as
\be
\label{initialization}
\mathcal{D}=[ -2^{6-d}, 2^{6-d} ]^d,\quad
V(\bx)=\fl{|\bx|^2}{2},\quad \psi_0(\bx)=\fl{1}{\sqrt[4]{\pi^{d}}}e^{-\fl{|\bx|^2}{2}},\quad
 \bx\in\mathcal{D}\;\; {\rm with}\;\; d=2\;\; {\rm or}\;\; 3.
\ee
Furthermore, the dipole orientations are chosen as $\bn=\bm=(0,0,1)^T$ in 3D and
$\bn_\perp=\bm_\perp=(1,0)^T$ in 2D, respectively.

 %test of accuracy in spatial direction
\begin{table}[h!]
\tabcolsep 12pt  \caption{
Spatial errors (upper parts)  $e_{\psi}^{h,\, \Dt t_0}(t)$ and temporal errors (lower parts)  $e_{\psi}^{h_0,\, \Dt t}(t)$  at $t=0.28$ for the dynamics of the 3D GPE (\ref{DipGPE1}) with different $\beta$ and $\lambda=\fl{\beta}{2}$.}
\label{spatial_temporal_3d_dy}
\begin{center}\vspace{-1.5em}
\def\temptablewidth{1\textwidth}
{\rule{\temptablewidth}{1pt}}
\begin{tabularx}{\temptablewidth}{@{\extracolsep{\fill}}p{1.25cm}cccccc}
$e_{\psi}^{h,\, \Dt t_0}(t)$ & $h=1/2$ & $h = 1/4$ & $h = 1/8$  & $h = 1/16$  \\ \hline
$\bt = 2$  &  3.999E-03	& 1.612E-05 & 1.601E-11  & 3.049E-12\\
$\bt = 10$ &  1.773E-02	& 2.581E-04 & 8.899E-09	 & 3.133E-12\\
$\bt = 50$ &  8.074E-02 & 8.186E-03 & 2.460E-05  & 2.304E-11\\[0.2cm]
%{\rule{\temptablewidth}{1pt}}
%\begin{tabularx}{\temptablewidth}{@{\extracolsep{\fill}}p{1.35cm}cccccc}
\hline\hline
$e_{\psi}^{h_0,\, \Dt t}(t)$ & $\Dt t=0.008$ & $\Dt t/2$ & $\Dt t/4$ & $\Dt t/8$  \\ \hline
$\bt = 2$  & 2.983E-06   &  7.454E-07 & 1.860E-07	 & 4.615E-08	 \\
 \text{rate} &  -- & 2.001  &  2.003  &   2.011  \\
$\bt = 10$ &  8.151E-06 &  2.036E-06  &  5.081E-07	 &   1.261E-07	  \\
 \text{rate} &  -- & 2.001 &   2.003 &   2.011 \\
$\bt = 50$ & 8.427E-05	     & 2.105E-05	 & 5.251E-06& 1.303E-06 \\
 \text{rate} & --& 2.001 &    2.003 &    2.011  \\
\end{tabularx}
{\rule{\temptablewidth}{1pt}}
\end{center}
\end{table}

\medskip

 \begin{exmp}{\sl Numerical accuracy verification in 3D.}\end{exmp}
Here d=3 and the ``exact" solution $\psi(\bx,t)$ is obtained numerically via the {\sl TS-NUFFT}
method on domain $\mathcal{D}$ with very small mesh size $h=h_0:=\fl{1}{16}$ and time step $\Dt t=\Dt t_0:=10^{-4}$.
Table \ref{spatial_temporal_3d_dy} lists the spatial discretization errors $e_{\psi}^{h,\, \Dt t_0}(t)$
and  the temporal discretization errors $e_{\psi}^{h_0,\, \Dt t}(t)$ as well as the convergence rate
at time $t=0.28$ with different mesh size  $h$ and different time step $\Dt t$, for different $\beta$ and $\lambda=\fl{\beta}{2}$.

\begin{exmp}  {\sl  Numerical accuracy verification in 2D.}\end{exmp}
 Here d=2 and the ``exact" solution $\psi(\bx,t)$ is obtained numerically via the {\sl TS-NUFFT}
method on domain $\mathcal{D}$ with very small mesh size $h=h_0:=\fl{1}{32}$ and time step
$\Dt t=\Dt t_0:=10^{-4}$.
Table \ref{spatial_temporal_2d_dy} shows the spatial discretization errors $e_{\psi}^{h,\, \Dt t_0}(t)$
and  the temporal discretization errors $e_{\psi}^{h_0,\, \Dt t}(t)$ as well as the convergence rate
at time $t=1.0$ with different mesh size  $h$ and different time step $\Dt t$,  for different $\beta$ and $\lambda=\fl{\beta}{20}$.

 %test of accuracy in spatial direction
 \begin{table}[h!]
\tabcolsep 0pt \caption{Spatial errors (upper parts)  $e_{\psi}^{h,\, \Dt t_0}(t)$  and temporal errors (lower parts)  $e_{\psi}^{h_0,\, \Dt t}(t)$  at $t=1.0$ for the dynamics of the 2D GPE (\ref{DipGPE1}) with different $\beta$ and $\lambda=\fl{\beta}{20}$.}
\label{spatial_temporal_2d_dy}
\begin{center}\vspace{-1.5em}
\def\temptablewidth{1\textwidth}
{\rule{\temptablewidth}{1pt}}
\begin{tabularx}{\temptablewidth}{@{\extracolsep{\fill}}p{1.35cm}cccccc}
$e_{\psi}^{h,\, \Dt t_0}(t)$ & $h=1/2$ & $h= 1/4$ & $h = 1/8$  &  $ h = 1/16$  \\ \hline
$\bt = 2$   &  5.715E-05 &  6.193E-11  &  1.120E-11	 &  1.124E-11	 \\
$\bt = 10$  &  1.894E-03 &  6.616E-08  &  1.354E-11 	 &   1.679E-11	  \\
$\bt = 50$  &  7.265E-02 &  2.987E-04  &  4.987E-10    &   2.852E-11\\[0.2cm]
\hline\hline
%{ \rule{\temptablewidth}{1pt} }
%\end{tabularx}
%{\rule{\temptablewidth}{1pt}}
%\begin{tabularx}{\temptablewidth}{@{\extracolsep{\fill}}p{1.25cm}cccccc}

$e_{\psi}^{h_0,\, \Dt t}(t)$ & $\Delta t =0.01$ & $\Delta t /2$ & $\Delta t /4$ & $\Delta t /8$  \\ \hline
$\bt = 2$  & 9.011E-06  &  2.252E-06   &  5.623E-07	 &  1.399E-07	 \\
\text{rate} &--& 2.001  &  2.002 &   2.007 \\	
$\bt = 10$ & 2.293E-05  &  5.728E-06   &  1.430E-06   	 &   3.558E-07\\
\text{rate} &--& 2.001 &   2.002 &    2.007\\	
$\bt = 50$ & 2.453E-04  &  6.122E-05	 &  1.528E-05       & 3.802E-06  \\
\text{rate} &--& 2.003  &  2.002  &   2.007 \\	
\end{tabularx}
{\rule{\temptablewidth}{1pt}}
\end{center}
\end{table}

\medskip
From Tabs.  \ref{spatial_temporal_3d_dy}-\ref{spatial_temporal_2d_dy}, we can see that the
{\sl TS-NUFFT} method (\ref{tssp2}) is  spectrally accurate in space and second order accurate in time
for computing the dynamics of dipolar BEC.

\subsection{Applications}
In this section,  we apply the TS-NUFFT method (\ref{tssp2}) to study some interesting phenomena, such as the
 dynamics of a BEC with time-dependent dipole orientations and the collapse and explosion of a dipolar
 BEC with attractive interaction and damping terms.

\begin{exmp}\label{dy_app_tune} {\sl  Dynamics of a BEC with rotating dipole orientations.}\end{exmp}

Here d=3 and we consider the GPE (\ref{DipGPE1}) without damping term, i.e., $f(\rho)\equiv0$.
The trapping potential is chosen as $V(\bx)=\fl{|\bx|^2}{2}$ and
the initial data in (\ref{ini-new}) is chosen as
$\psi_0(\bx)=\phi_{\rm gs}(\bx)$, where $\phi_{\rm gs}(\bx)$ is
  the ground state of the GPE (\ref{DipGPE1}) with $f(\rho)\equiv0$ and $\bn=\bm=(0,0,1)^T,$ $\beta=103.58$
  and $\lambda=82.864$, which is computed numerically via the numerical method
   presented in the previous section. The computational domain and mesh size are chosen as
  $\mathcal{D}=[-8, 8]^3$ and $h_x=h_y=h_z=\fl{1}{8}$, respectively.
Then  we  tune the dipole orientations as
\be\label{dip_rot}
\bn(t)=\left( \sin\fl{t}{5}, 0, \cos\fl{t}{5} \right)^{T}, \qquad t\ge0,
\ee
and study the dynamics of the BEC in two cases:

\begin{itemize}

\item Case 1. tune the dipole orientation as in \eqref{dip_rot} and keep all the other parameters unchanged.

\item Case 2. tune the dipole orientation as in \eqref{dip_rot}, perturb the trapping potential by setting $\gamma_x=2$ and keep all the other parameters unchanged.
\end{itemize}

\begin{figure}[h!]
\centerline{
\psfig{figure=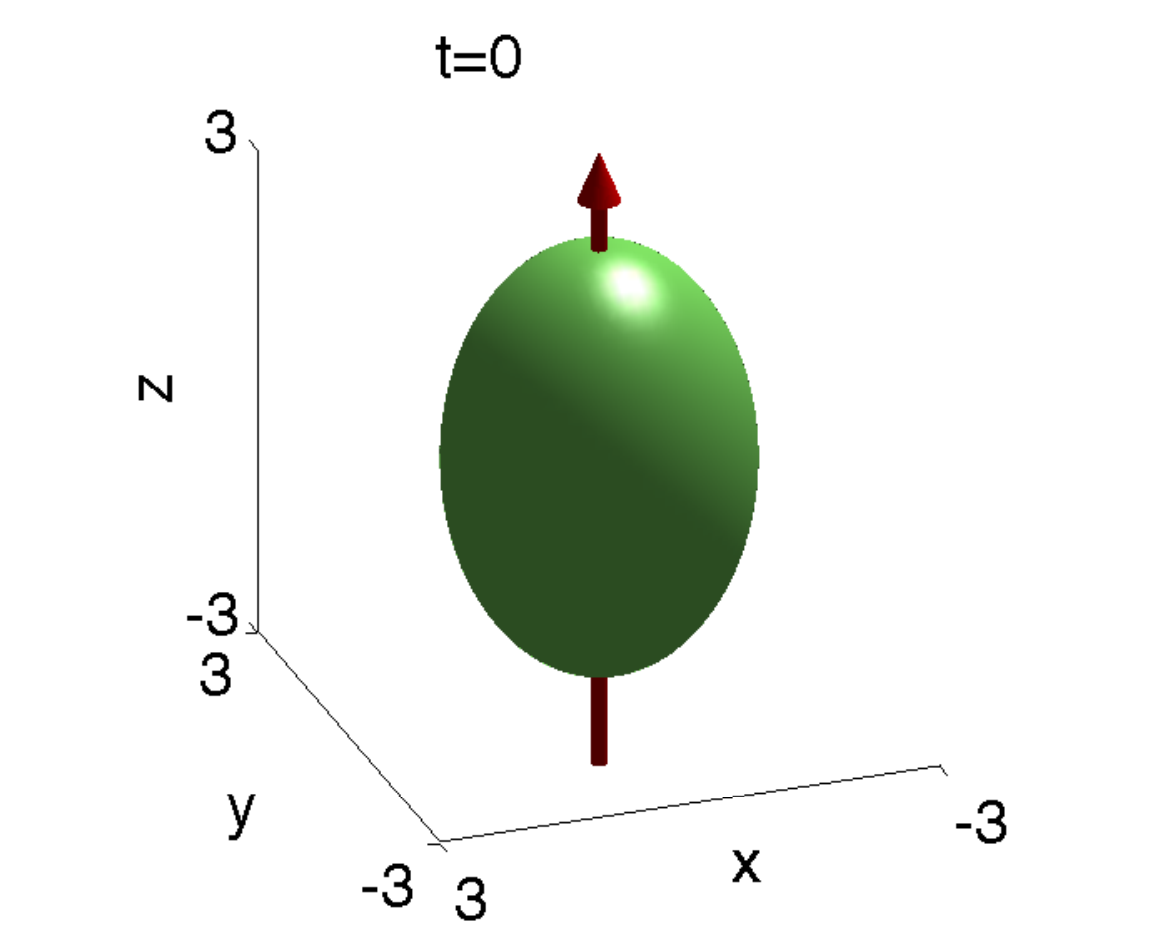,height=3.8cm,width=4.4cm,angle=0}\quad
\psfig{figure=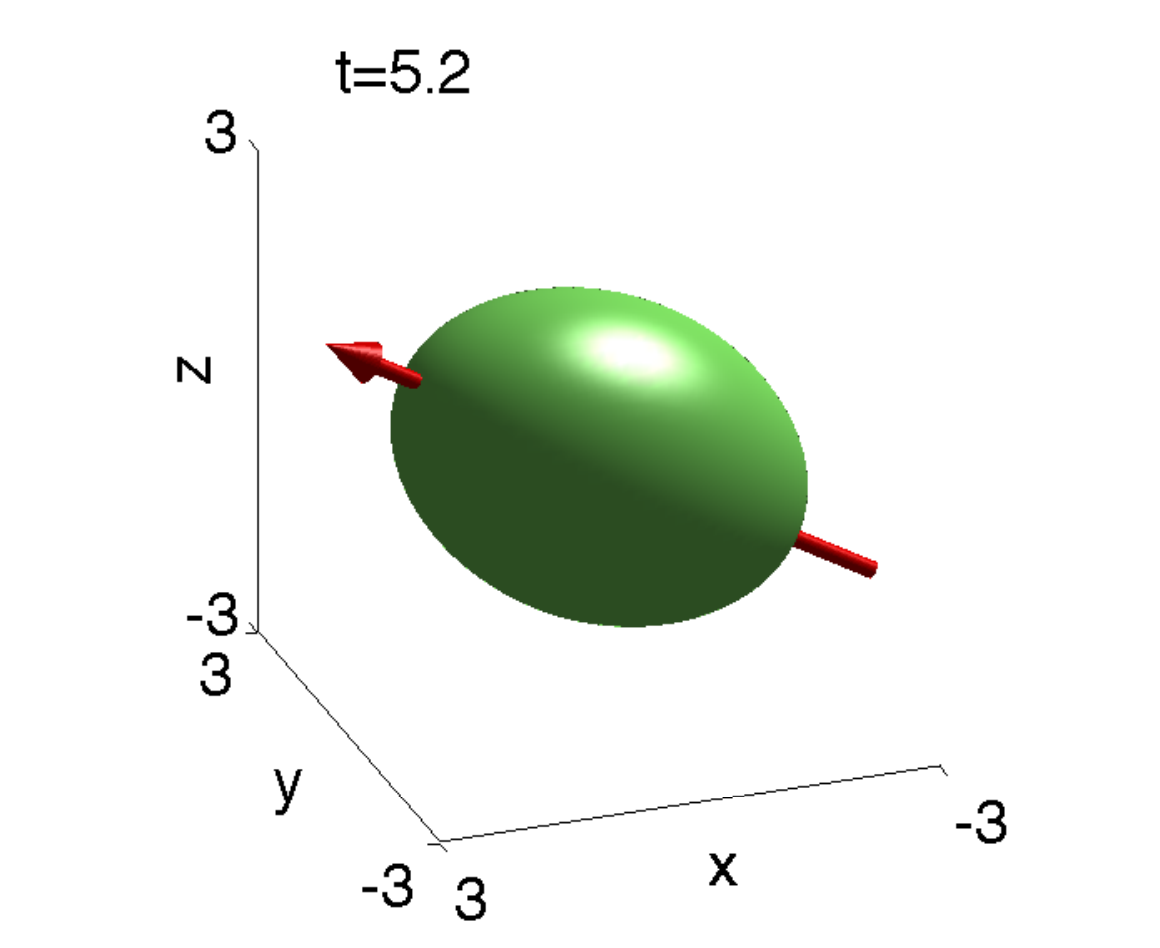,height=3.8cm,width=4.4cm,angle=0}\quad
\psfig{figure=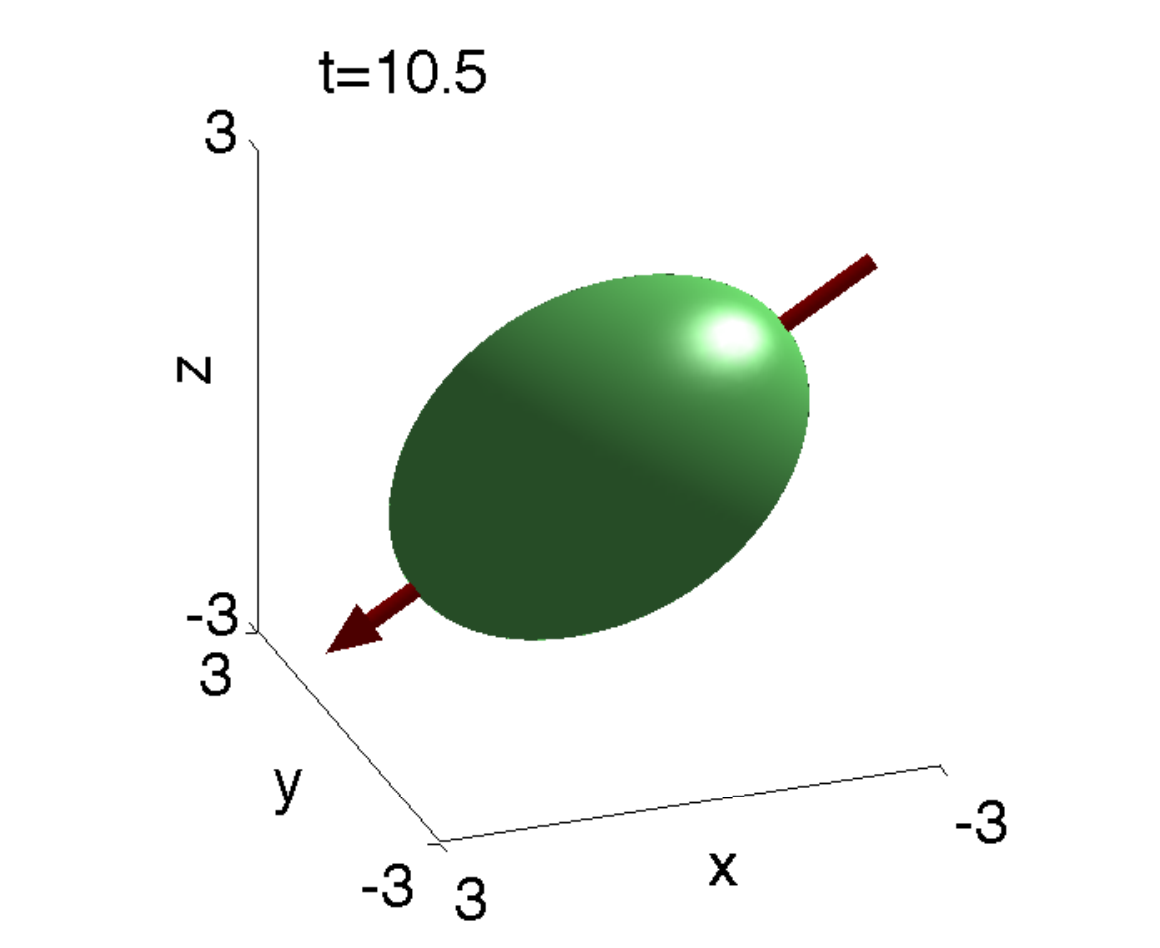,height=3.8cm,width=4.4cm,angle=0}
}
\centerline{
\psfig{figure=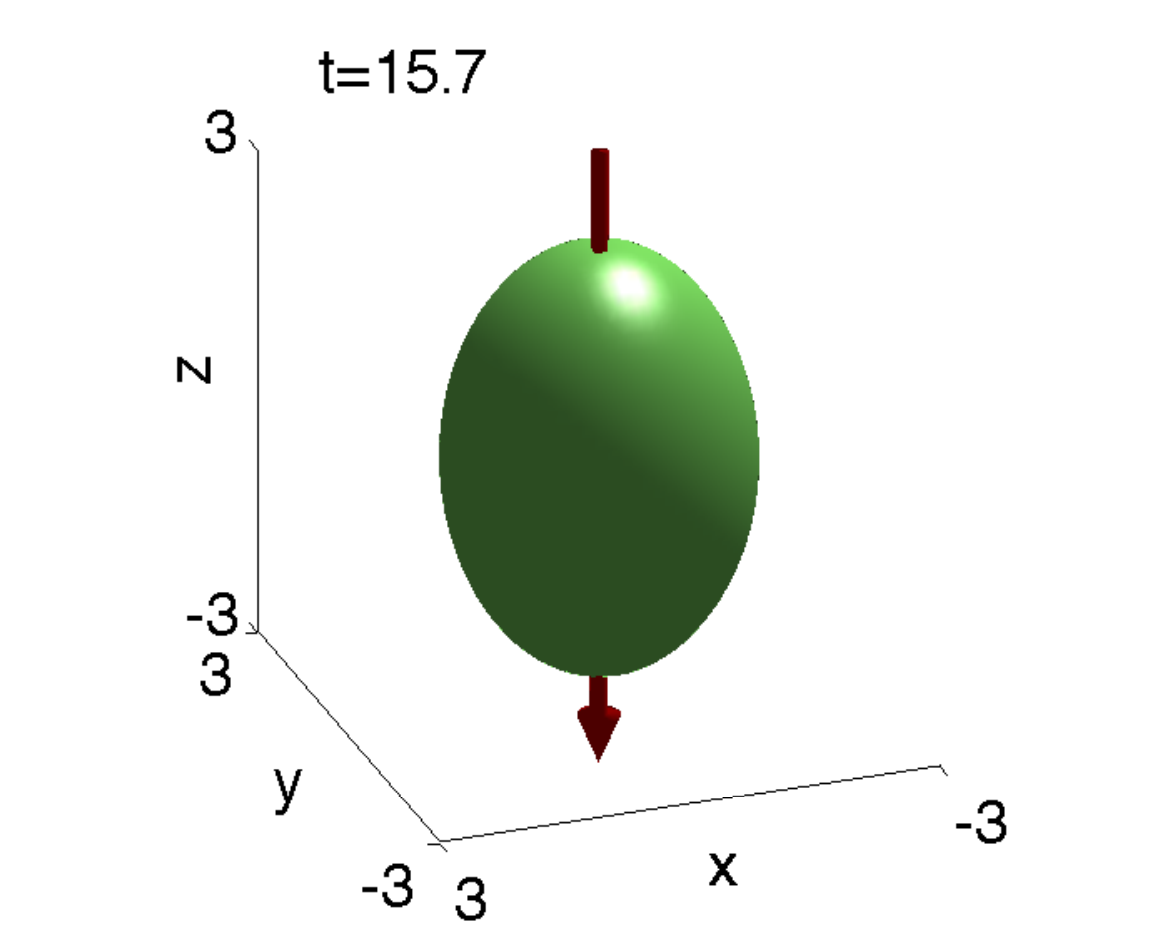,height=3.8cm,width=4.4cm,angle=0}\quad
\psfig{figure=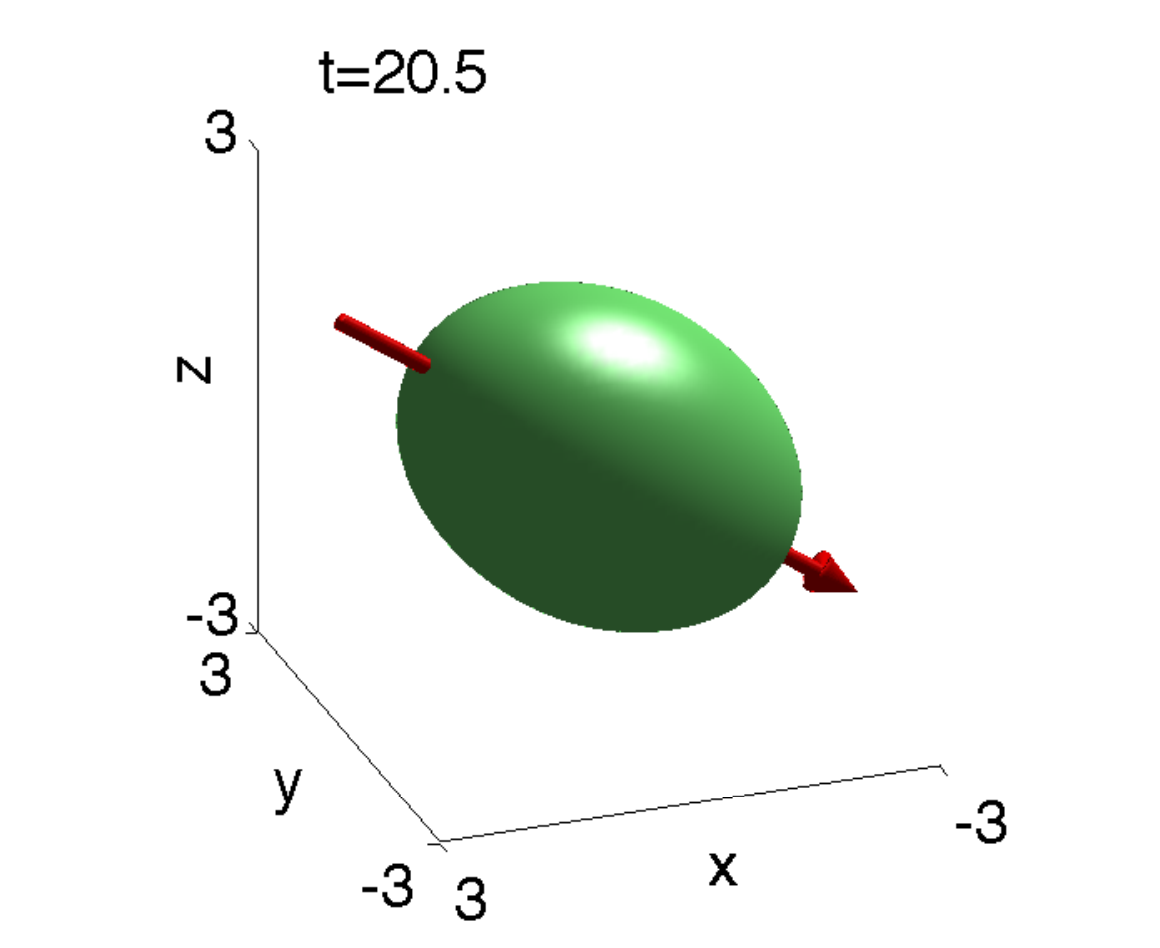,height=3.8cm,width=4.4cm,angle=0}\quad
\psfig{figure=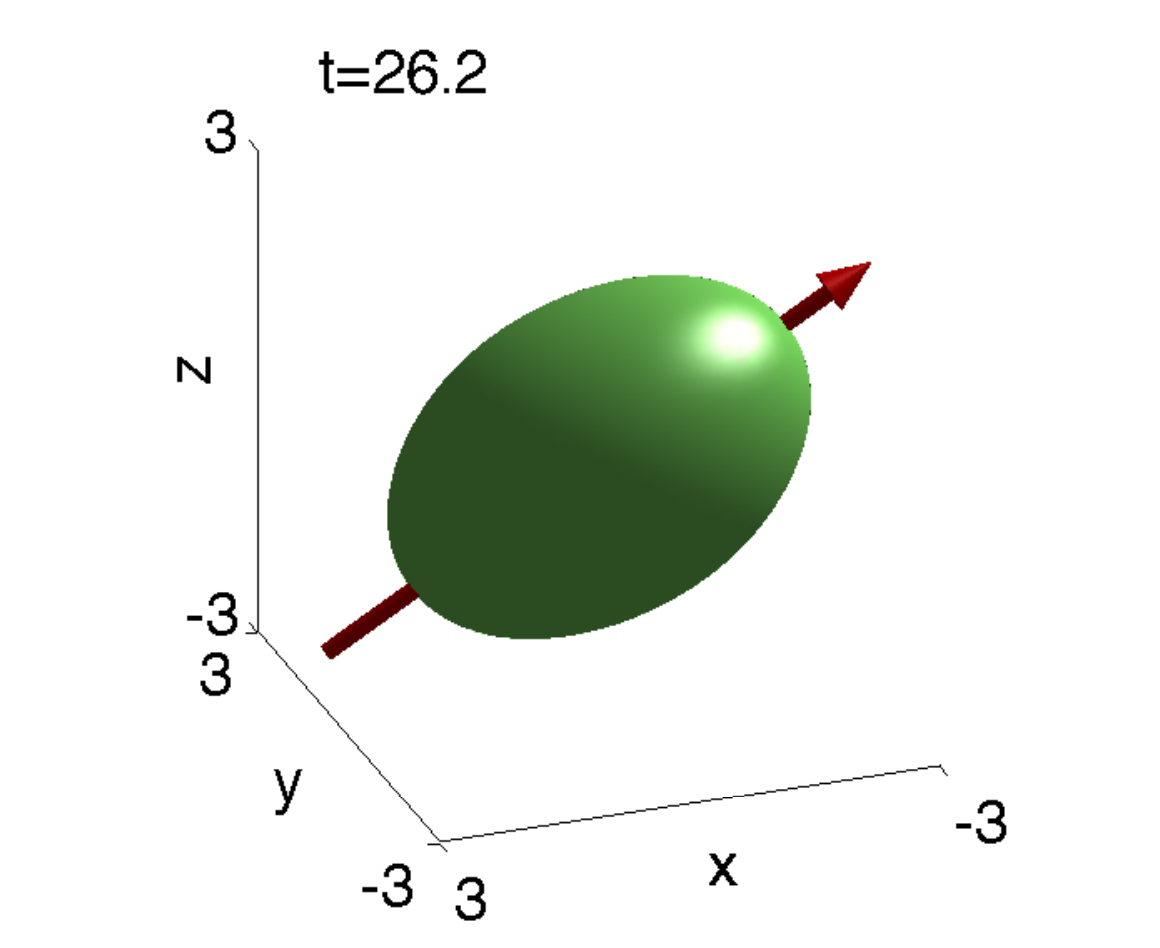,height=3.8cm,width=4.4cm,angle=0}
}
\caption{Isosurface plots of the density function $\rho(\bx,t)=|\psi(\bx,t)|^2=0.01$
and the dipole axis $\bn(t)$ (red arrow)  at different times for case 1 in the
example \ref{dy_app_tune}.}
\label{fig:dens-ex1}
\end{figure}

\begin{figure}[h!]
\centerline{
\psfig{figure=./figures/dynamics/dens10.pdf,height=3.8cm,width=4.4cm,angle=0}\quad
\psfig{figure=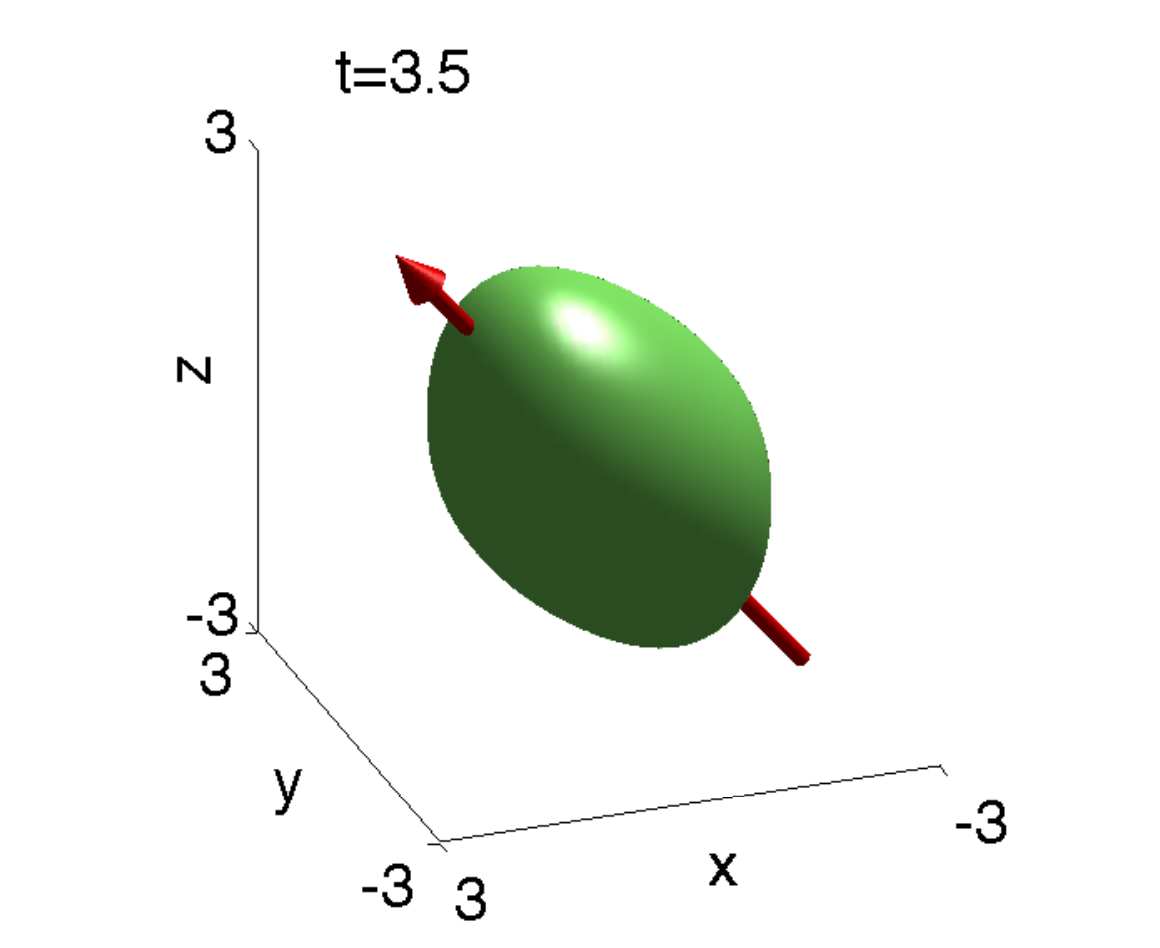,height=3.8cm,width=4.4cm,angle=0}\quad
\psfig{figure=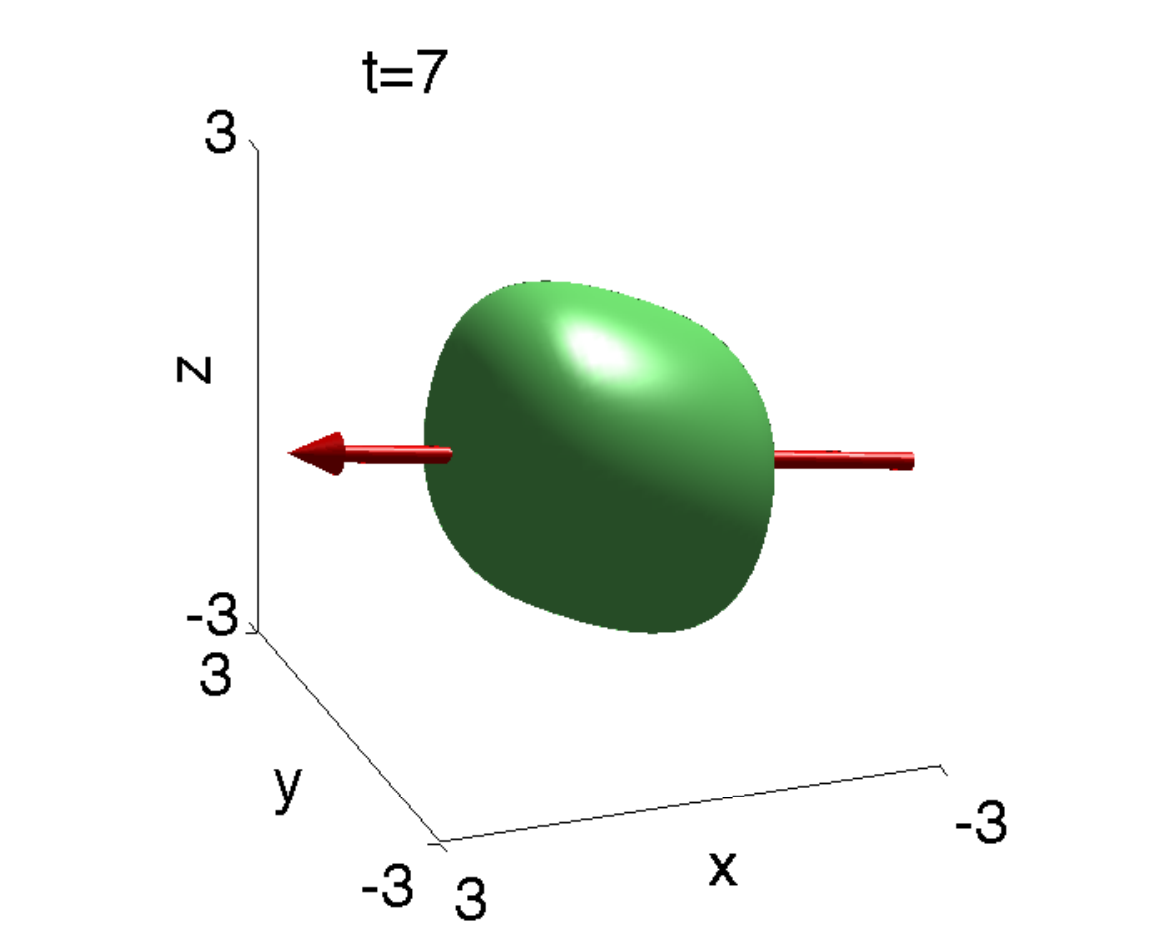,height=3.8cm,width=4.4cm,angle=0}
}
\centerline{
\psfig{figure=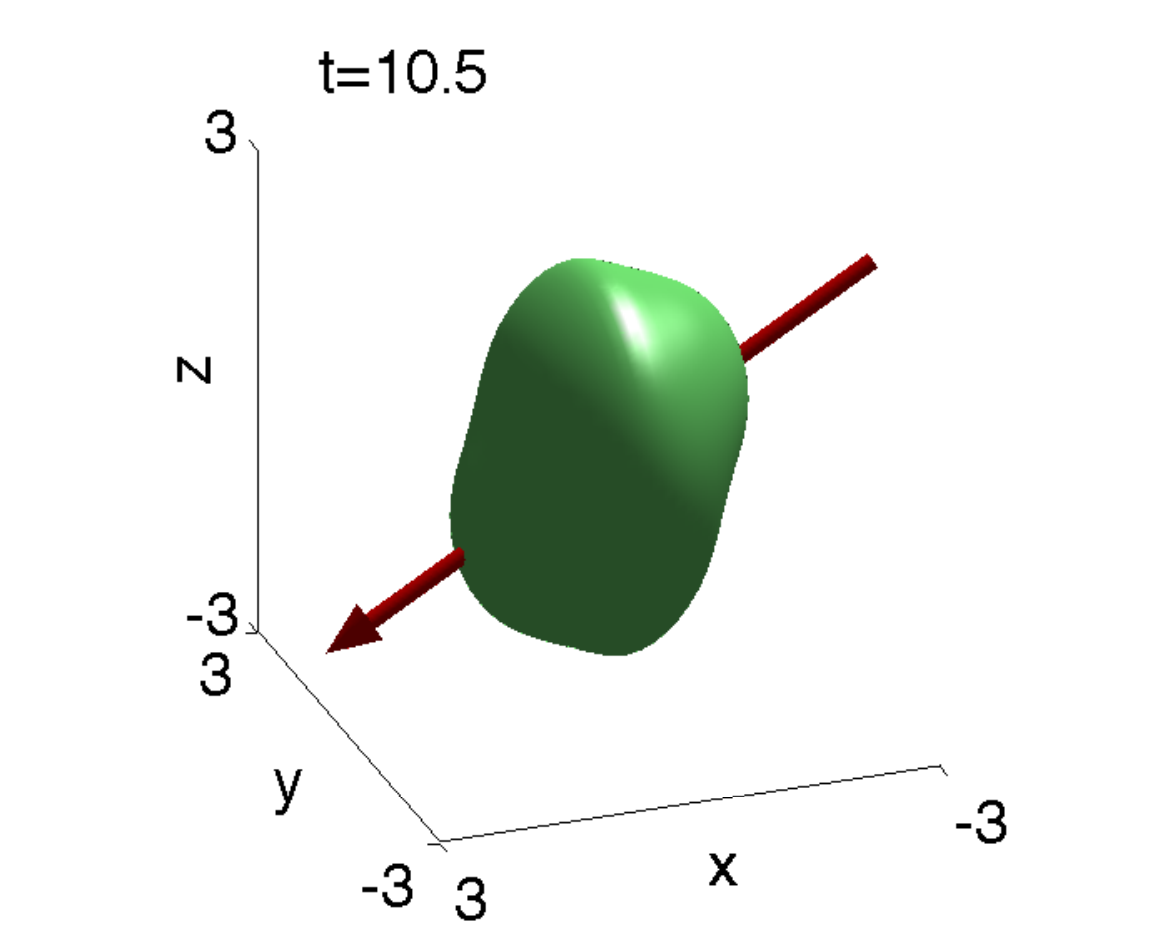,height=3.8cm,width=4.4cm,angle=0}\quad
\psfig{figure=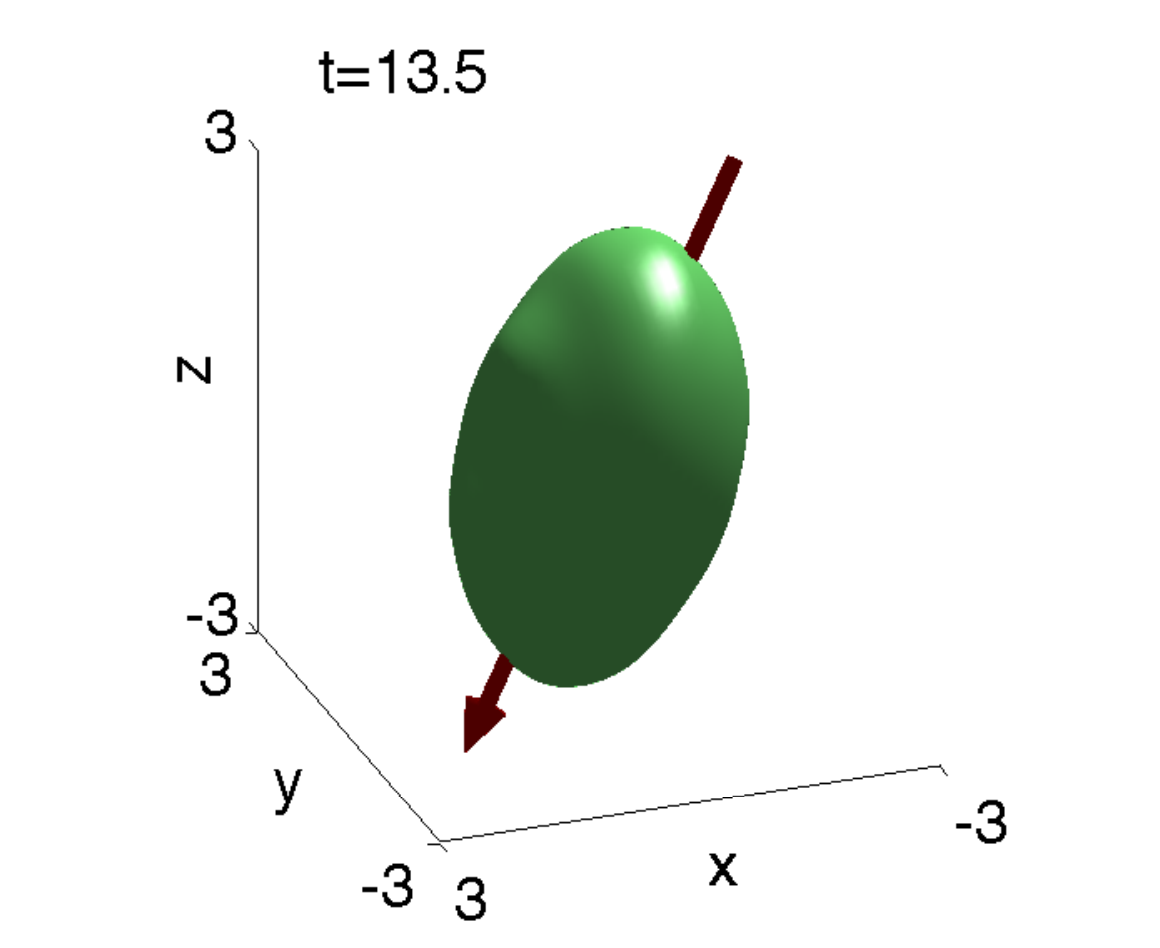,height=3.8cm,width=4.4cm,angle=0}\quad
\psfig{figure=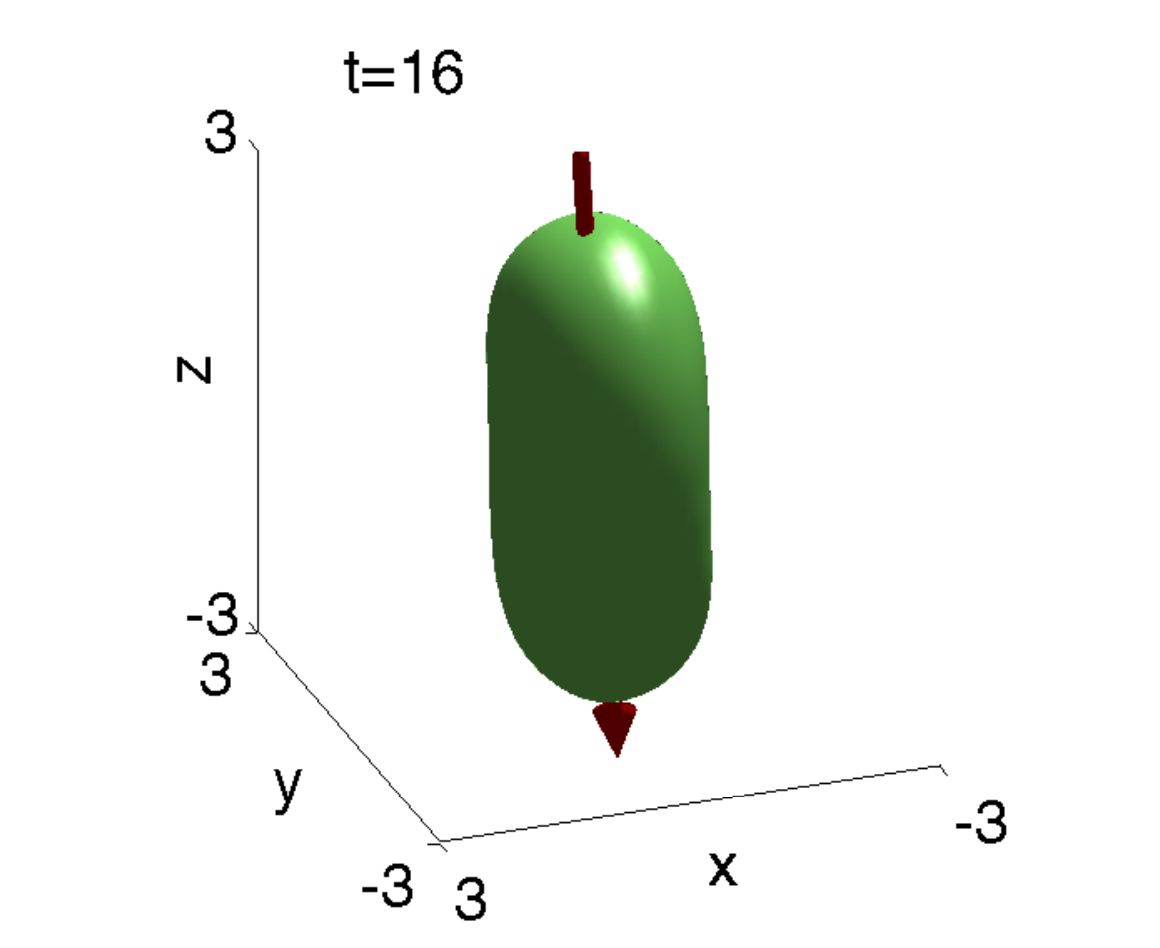,height=3.8cm,width=4.4cm,angle=0}
}
\caption{Isosurface plots of the density function $\rho(\bx,t)=|\psi(\bx,t)|^2=0.01$
and the dipole axis $\bn(t)$ (red arrow)  at different times for  case 2 in the example \ref{dy_app_tune}.}
\label{fig:dens-ex1-Ext}
\end{figure}

Figure \ref{fig:dens-ex1} shows the isosurface of the density function $\rho(\bx,t)=|\psi(\bx,t)|^2=0.01$ at different times for case 1, while Figure \ref{fig:dens-ex1-Ext} depicts the isosurface evolution for case 2. From Figs.  \ref{fig:dens-ex1}-\ref{fig:dens-ex1-Ext}, we could have the following conclusions:
(i). The density of the condensate will rotate along with the rotation of  the dipole axis.
(ii). For Case 1 where the trapping potential is isotropic,
 the shape of the density profile seems to be unchanged
during the dynamics,  and it seems to keep the same symmetric structure with
respect to the dipole orientation.  However, this phenomena does not occur
in Case 2 where the trapping potential is anisotropic.

%(iii).  From additional numerical experiments, we can conjecture that:  if the initial data
%$\psi_0(\bx)$ is chosen as a ground state $\psi_0(\bx)$ of GPE with isotropic potentials, i.e.,
%$\gamma_x=\gamma_y=\gamma_z$ and fixed dipole axis $\bn_0$,  then by simply rotating the dipole axis
%via $\bn(t)={\bf B}(t) \bn_0$ where $B(t)$ is the rotating matrix in 3D,  the analytical solution of
%the GPE (\ref{DipGPE1}) would reads as  $\psi_0({\bf B}(t)\bx)$ if additionally the following holds true
%\be
%\Big( \fl{d {\bf B}}{dt}\_{t=0}\, \bx \Big) \cdot \nabla\psi_0=0.
%\ee
%

\begin{exmp}\label{dy_app_damp}  {\sl  Collapse and explosion of a dipolar BEC with damping effect in 3D.} \end{exmp}

\begin{figure}[h!]
\centerline{
\psfig{figure=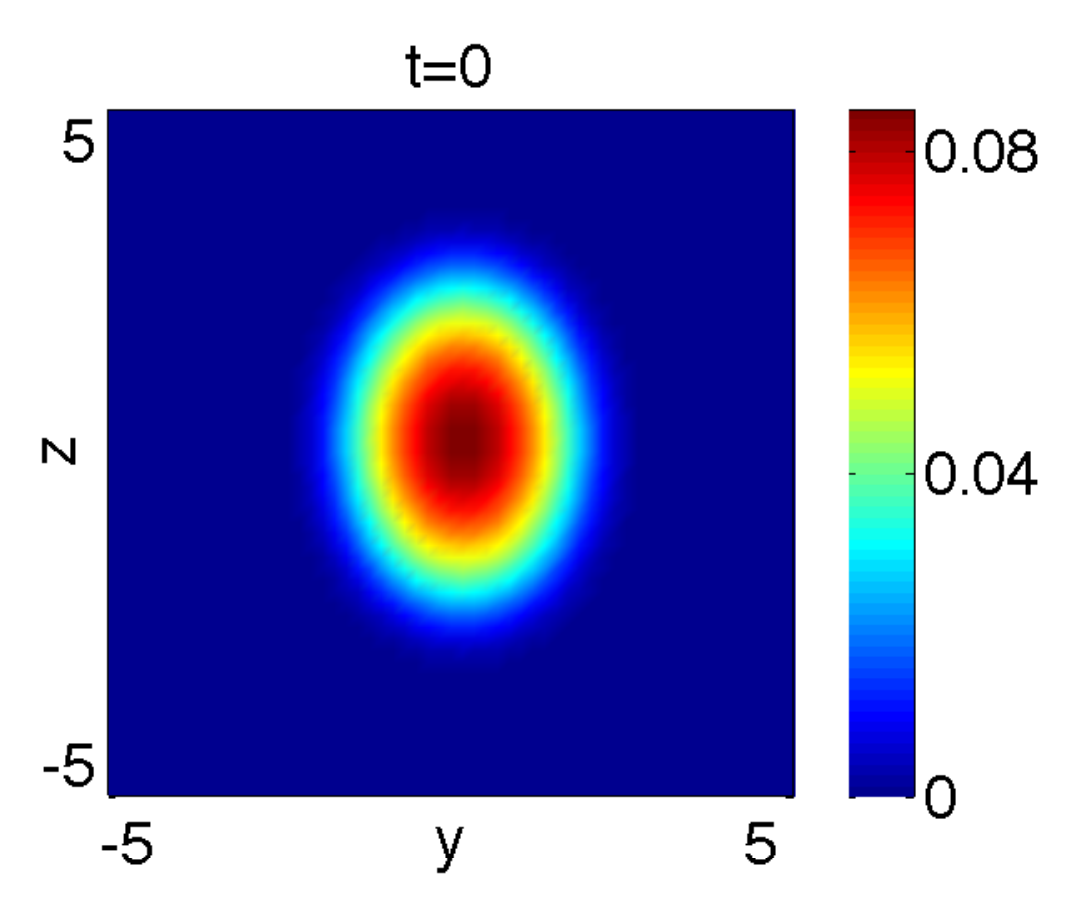,height=3.8cm,width=4.4cm,angle=0}\;\;\;
\psfig{figure=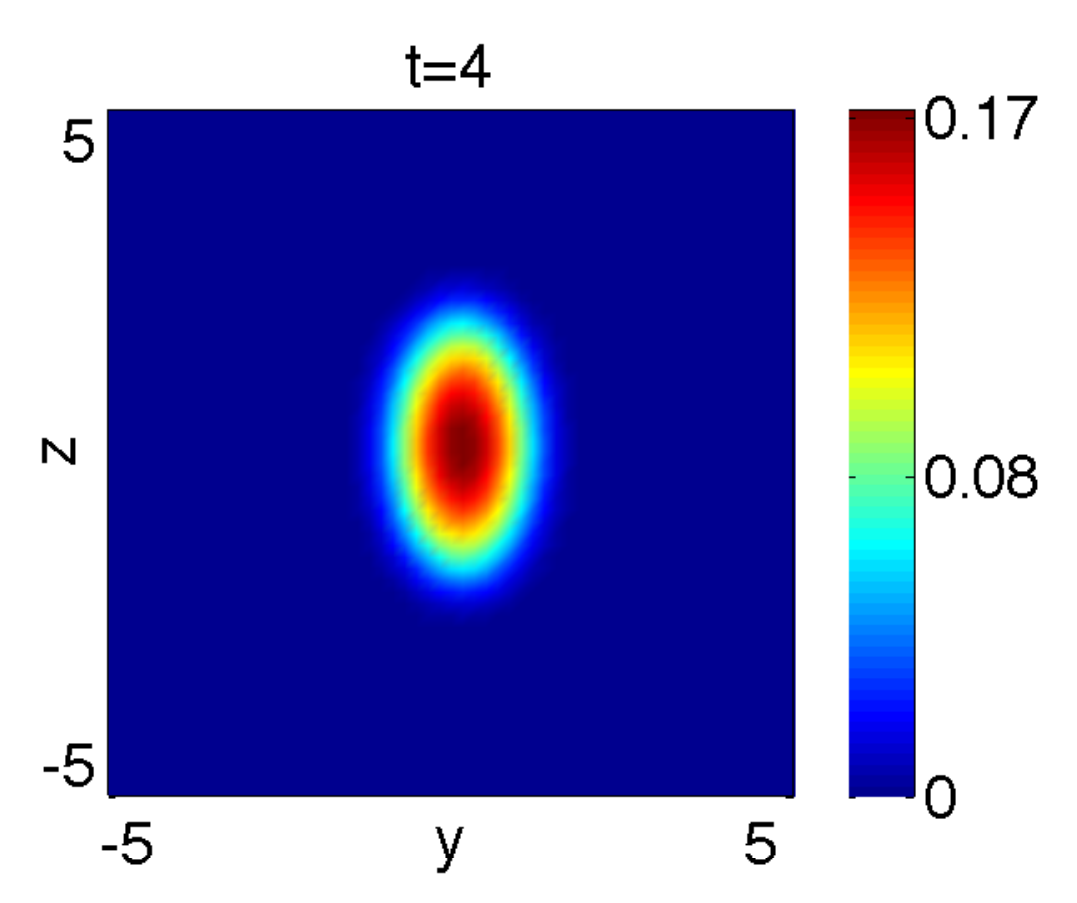,height=3.8cm,width=4.4cm,angle=0}\;\;\;
\psfig{figure=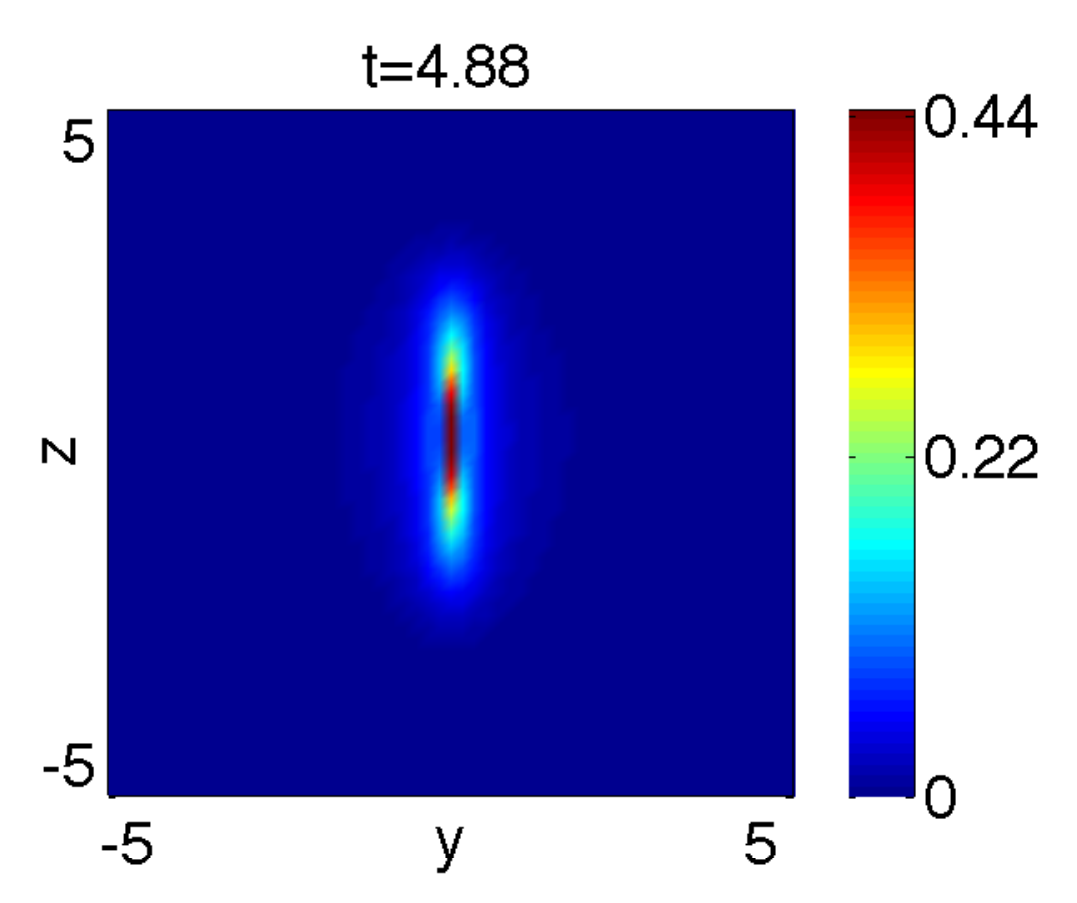,height=3.8cm,width=4.4cm,angle=0}
}
\centerline{
\psfig{figure=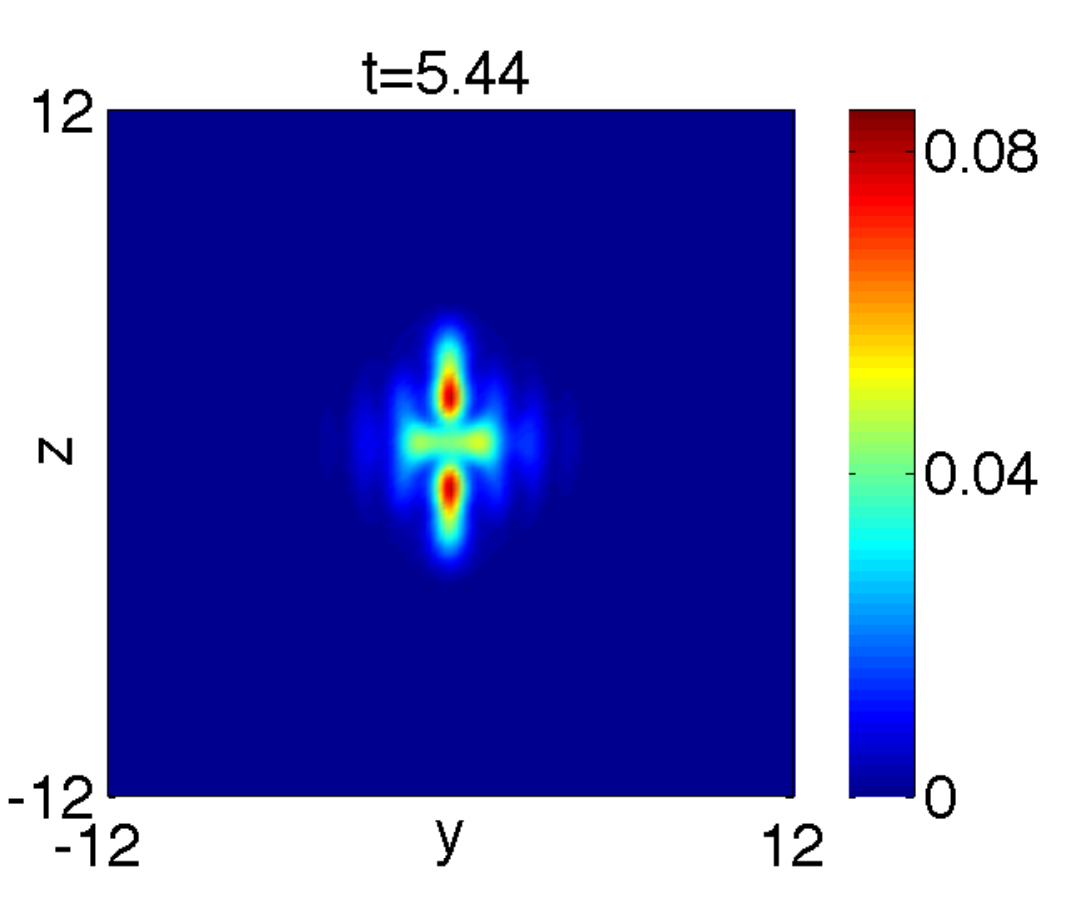,height=3.8cm,width=4.4cm,angle=0}\;\;\;
\psfig{figure=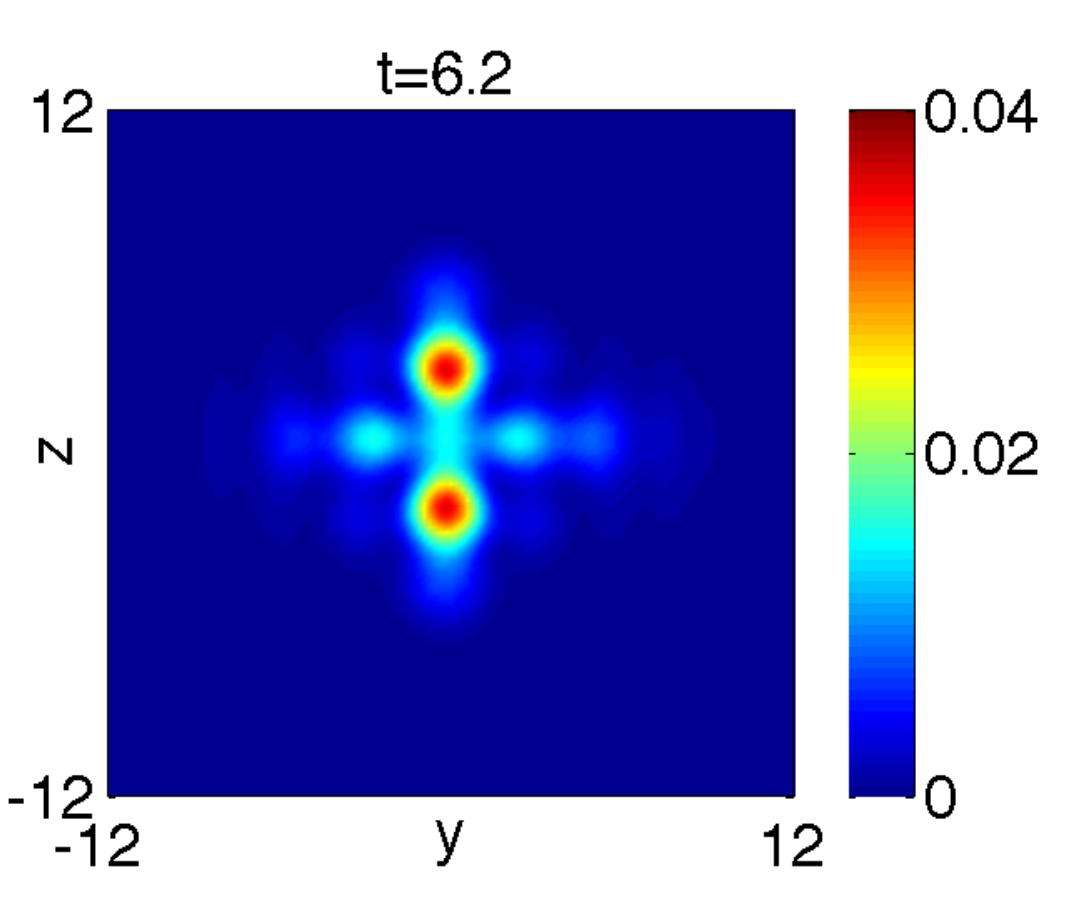,height=3.8cm,width=4.4cm,angle=0}\;\;\;
\psfig{figure=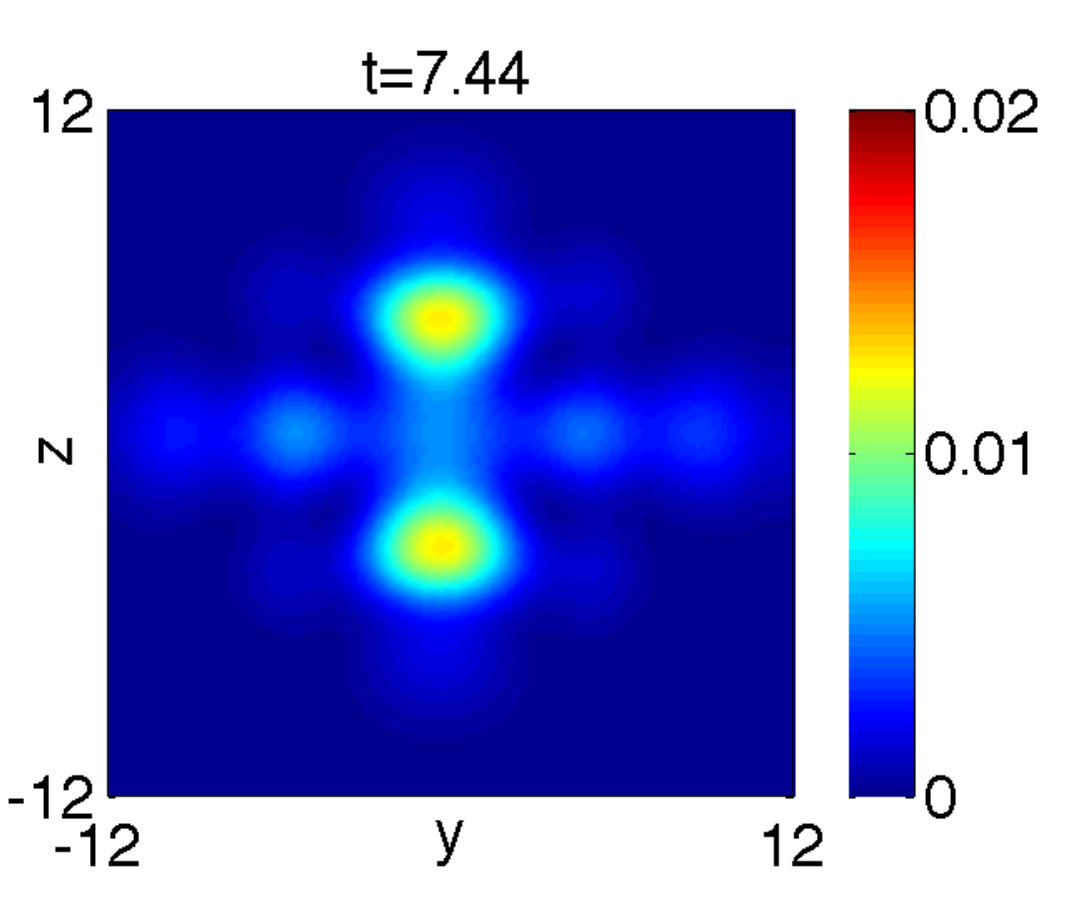,height=3.8cm,width=4.4cm,angle=0}
}
\caption{Contour plots of the column density  $\rho_c^x(y,z,t)$  at different times for the example \ref{dy_app_damp}.}
\label{fig:column-dens-ex3}
\end{figure}

\begin{figure}[h!]
\centerline{
\psfig{figure=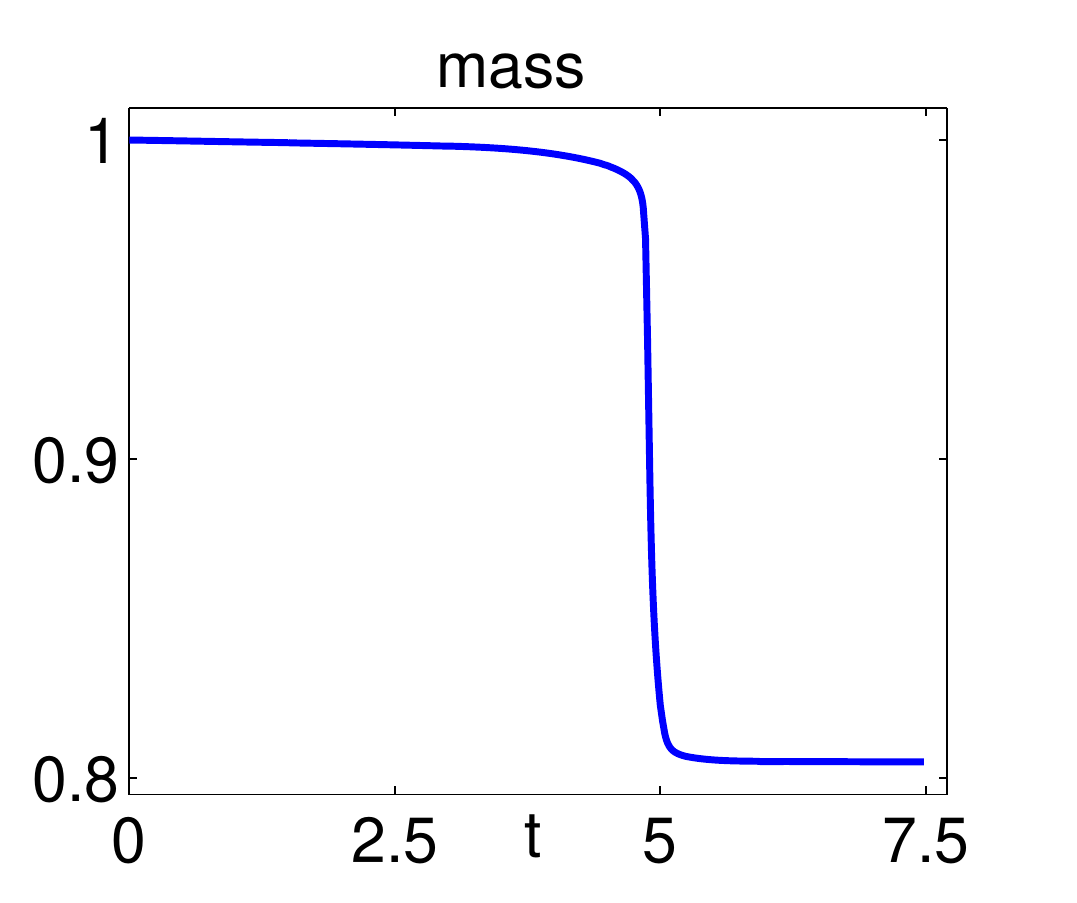,height=4.4cm,width=8cm,angle=0}
}
\caption{Evolution of the mass for the example \ref{dy_app_damp}.}
\label{fig:mass-ex3}
\end{figure}

In this case,  the trapping potential $V(\bx)$ and the constants $\lambda$ and  $\beta$  are set to
be time dependent and are chosen according to the parameters used in the physical experiment
\cite{Lah2008,LMSLP2009} (in dimensionless form) as follows:
\be
V(\bx,t)=\left\{
\begin{array}{ll}
(\gamma_x^2x^2+\gamma_y^2y^2+\gamma_z^2z^2)/2,&\quad t\in[0,\; 4+t_{\rm hold}],\\[0.5em]
0, &\quad {\rm otherwise,}
\end{array}
\right.
\ee
with $\gamma_x=1.65$, $\gamma_y=1$, $\gamma_z=1.325$,
\bea
\lambda(t)=
\left\{\begin{array}{ll}
82.864, &\quad t\in[0,\; 5.6+t_{\rm hold}],\\[0.5em]
0, & \quad {\rm otherwise},
\end{array}
\right.
\eea
\bea
\beta(t)=761.102\left\{
\begin{array}{ll}
1+\frac{28}{75t-280}, &\quad t\in[0,\;3.2], \\[0.5em]
0.3,	&\quad t\in[3.2,\;3.6],	\\[0.5em]
1-\frac{1}{b(t-3.6)}, &\quad t\in[3.6,\;4.8], \\[0.5em]
38.8066, &\quad t\in [4.8,\; 5.6+t_{\rm hold}], \\[0.5em]
0, &\quad {\rm otherwise},
\end{array}
\right.
\eea
where
\bea
b(t)=\fl{1}{133}\left\{
\begin{array}{ll}
-125t-25e^{-5t}+215, &\quad t\in[0,\;0.4],\\[0.5em]
25(1-e^{-2})e^{-6.25t+2.5}+140,&\quad t\in[0.4,\;1.2].
\end{array}
\right.
\eea
Here, $t_{\rm hold}$  is the hold time for the collapse, which is chosen as
 $t_{\rm hold }=0.2$.
Moreover, we let $\bn=(0, 0, 1)^T$,
$\sigma=1$ and chose the damping term  as
$f(\rho)=\delta \rho^2$ with $\delta=3.512$, i.e., we chose the cubic nonlinearity and
study the case of three-body loss of the particles.
The initial data in (\ref{ini-new}) is chosen as
$\psi_0(\bx)=\phi_{\rm gs}(\bx)$, where $\phi_{\rm gs}(\bx)$ is
  the ground state of the GPE (\ref{DipGPE1}) with $f(\rho)\equiv0$, $\bn=\bm=\bn(0)$, $\beta=\beta(0)$
  $\lambda=\lambda(0)$ and $V(\bx)=V(\bx,0)$, which is computed numerically via the numerical method
   presented in the previous section.
The computational domain and mesh size are chosen as
 $\mathcal{D}=[-24, 24]^3$ and $h_x=h_y=h_z=\fl{3}{16}$, respectively.
Figures \ref{fig:column-dens-ex3} and \ref{fig:mass-ex3}  show the contour plot of the column density
\[
\rho_c^x(y,z,t)=\int_{L_x}^{R_x}|\psi(\bx,t)|^2dx,
\]
and the evolution of the total mass, respectively.

 From Figs. 4 and 5, we can conclude that:
(i). The total mass is lost during the dynamics, especially during
a very short period near $t=5$ (cf. Fig. 5). (ii). Although the BEC is released from the trap (i.e., the trapping potential is turned off) at time $t=4.2$, the atoms in the BEC still move inward in the $x$-$y$ plane. (iii).
The density is first enlongated   along the dipole
orientation, then the collapse of the BEC happens very quickly, and ``clover" pattern of the density profile
is created.  (iv). The ``Leafs" are then ejected outward.
All these results agree with those in the experiments \cite{Lah2008,LMSLP2009}.

%%%%%%%%%%%%%%%%%%%%%%%%%%%%%%%%%%%%%%%%%%%%%%
%%%%%%%%%%%%%%%%%%%%%%%%%%%%%%%%%%%%%%%%%%%%%%
\section{Conclusion}
We proposed efficient and accurate numerical methods for computing the ground state
and dynamics of the dipolar Bose-Einstein condensates by integrating
a newly developed dipole-dipole interaction (DDI) solver via
the non-uniform fast Fourier transform (NUFFT) algorithm \cite{BaoJiangLeslie}
with existing numerical methods.
The NUFFT based DDI solver removes naturally the singularity of the DDI
at the origin by adopting the spherical/polar coordinates in the Fourier space,
thus achieves spectral accuracy and simultaneously maintains high efficiency by
appropriately combining the advantages of the NUFFT and FFT.
Efficient and accurate numerical methods were then presented to compute the ground state and dynamics of the dipolar BEC with a DDI by integrating
the normalized gradient flow with the backward Euler Fourier pseudospectral discretization and time-splitting Fourier pseudospectral method, respectively,
together with NUFFT based DDI solver. Extensive numerical comparisons with  existing methods were carried out to compute the DDI, ground states and dynamics of the dipolar BEC. Numerical results showed that our new methods outperformed other existing methods in terms of both accuracy and efficiency,
especially when the computational domain is chosen smaller and/or the solution is anisotropic.

\

\section*{Acknowledgements}
We acknowledge support from
the Ministry of Education of Singapore grant R-146-000-196-112 (W. Bao),
the French ANR-12-MONU-0007-02 BECASIM (Q. Tang)
and  the Austrian Science Foundation (FWF) under grant No. F41 (project VICOM),
grant No. I830 (project LODIQUAS) and the Austrian Ministry of Science and
Research via its grant for the WPI (Q. Tang and Y. Zhang).  The
computation results presented have been achieved by using the Vienna Scientific Cluster.
This work was partially done while the authors were visiting Beijing Computational Science
Research Center in the summer of 2014, the Institute for Mathematical Sciences,
National University of Singapore, in 2015, and the Computer, Electrical and Mathematical
Sciences and Engineering Division, King Abdullah University of Science and Technology,
in 2014.

\end{document}